\documentstyle[11pt,aaspp4]{article}

\begin{document}

\title{Intermittent Features of the QSO Ly$\alpha$ Transmitted Flux:
Results from Hydrodynamic Cosmological Simulations}

\author{Long-Long Feng\altaffilmark{1,2},
       Jesus Pando\altaffilmark{3},
       and Li-Zhi Fang\altaffilmark{4}  }

\altaffiltext{1}{Center for Astrophysics, University of Science
and Technology of China, Hefei, Anhui 230026,P.R.China}
\altaffiltext{2}{National Astronomical Observatories, Chinese
Academy of Science, Chao-Yang District, Beijing, 100012, P.R.
China}
\altaffiltext{3}{Department of Physics, DePaul University, Chicago Il,
60614 }
\altaffiltext{4}{Department of Physics, University of Arizona,
Tucson, AZ 85721}

\begin{abstract}

It has been recently found that the local fluctuations of the  QSO's
Ly$\alpha$ absorption spectrum transmitted flux show spiky structures.
This implies that the mass fields of the intergalactic medium (IGM) is
intermittent. This feature cannot be explained by the clustering
evolution of cosmic mass field in the linear regimes and is also
difficult to incorporate into the hierarchical clustering scenario.
We calculate the structure functions and intermittent exponent of the IGM
and HI for full hydrodynamical simulation samples. The result shows
the intermittent features of the Ly$\alpha$ transmitted flux fluctuations
as well as the mass field of the IGM. We find that within the error bars
of current data, all the intermittent behavior of the simulation samples
are consistent with the observation. This result is different from our
earlier result (Pando et al 2002), which shows that the intermittent
behavior of samples generated by pseudo-hydro simulation cannot be fitted
with observed data. One difference between the pseudo-hydro and full
hydro simulations is in treating the dynamical relation between the IGM
(or HI) and dark matter fields. The former assumes that the IGM density
distribution traces the underlying dark matter point-by-point on scales
larger than the Jeans length in either the linear or nonlinear regimes.
However, hydrodynamic studies have found that a statistical discrepancy
between the IGM field and underlying dark matter in nonlinear regime is
possible. We find that the point-by-point correlation between the IGM
density perturbations and dark matter become weaker on comoving scales
less than 2 h$^{-1}$ Mpc (in LCDM model), which is larger than the IGM
Jeans length.

\end{abstract}

\section{Introduction}

In the past few years there has been tremendous progress in
understanding the evolution of cosmic structures on scales from
galaxies to about 100 h$^{-1}$ Mpc. This progress has occurred
mainly in two distinct areas: 1.) constraining cosmological
parameters such as the mass density, dark energy, power spectrum
of the initial perturbations of the cosmic mass field, etc., 2.) the
modeling of the origin of light-emitting objects to explain the
observed mass density profiles, velocity dispersions, mass-to-light
ratios, substructures, bias behavior, etc., of quasars, galaxies and
galaxy clusters.

These two distinct areas coincide with two different regimes in the
context of the dynamics of structure formation. The first one is based
mainly on the evolution of the cosmic mass field in the linear
approximation. The basic statistical tool is the power spectrum,
either observed or recovered by Gaussianization, of the weakly
non-linear field. The second regime is characterized by the dynamics of
collapsed massive halos. The formation and evolution of luminous clumps
can be modeled by the growing and merging of massive halos from
initially Gaussian perturbations of the mass and velocity fields.
The hierarchical clustering of halos is the most effective
scenario for modeling galaxy formation.

However, the clustering behavior of the mass and velocity fields
of Inter-Galactic Medium (IGM), i.e., the major component of
baryonic matter in the universe, is missing from these two regimes. It is
certain that the IGM field is not linear and Gaussian because the
distributions of QSO's Ly$\alpha$ forests from either observational
data or simulation samples (Bi 1993, Bi, Ge \& Fang 1995), have been
found to be non-Gaussian even on scales as large as a few h$^{-1}$ Mpc
(Pando \& Fang 1996, 1998a).
Neither is the evolution of the IGM field well modeled by the dynamics of
collapsing halos, as the QSO's Ly$\alpha$ forests
are generally not gravitationally confined (Fang et al 1996). The success
of modeling the QSO's Ly$\alpha$ forests via a lognormal model
of the IGM (Bi \& Davidsen 1997) directly indicates that the IGM field
probably does not (or has not yet) undergone a correlation hierarchy evolution.

The clustering behavior of the IGM mass field can play an
important role in the dynamical gap between linear power spectrum and collapsed
massive halos. For instance, it has recently been found that
the QSO Ly$\alpha$ transmitted flux fluctuations are intermittent
(Jamkhedkar, Zhan \& Fang 2000, Feng, Pando \& Fang 2001, Zhan, Jamkhedkar
\& Fang 2001).  A field is intermittent if the 2$n^{th}$ power of the local
differences grows faster than the $n^{th}$ power of the variance of these
differences (see Section 2.) This result cannot be found using the statistical
tools of the linear regime such as power spectrum.  At the same time, it is
unclear how to incorporate intermittent behavior with the hierarchical
clustering scenario. The intermittent features have also been
studied with simulated Ly$\alpha$ forests produced by a pseudo hydro
technique.  All the pseudo hydro samples for low density cold dark matter
(LCDM) and warm dark matter (WDM) models are found to be substantially
intermittent. Although the intermittent features of the simulation samples
were different from those of the observations (Pando et al 2002), the results
implies that intermittency is a common feature of non-linear evolution and
is potentially useful for model discrimination.

This motivated us to re-do the intermittent analysis with simulated
Ly$\alpha$ transmitted fluxes produced by a full hydrodynamical treatment.
This motivation was strengthened when we realized that the relation
between the IGM and dark matter fields in the non-linear regime is treated
differently in the pseudo-hydro and full hydro simulations. In the
pseudo-hydro simulation, the density of the IGM in each pixel is assigned
according to the dark matter density at that pixel. That is, the
pseudo-hydro simulation assumes that the density distribution of IGM traces
the underlying dark matter point-by-point on scales larger than the Jeans
length in either the linear or nonlinear regimes. This assumption looks
reasonable when one considers that the IGM is only a
passive substance in the cosmic mass field since its evolution is
dominated
by the gravity of dark matter.

However, hydrodynamic studies have shown that a passive substance in
the system does not always behave like the underlying dominant mass
field when non-linear evolution has taken place (for a review,
Shraiman \& Siggia 2001). The statistical properties of a passive
substance can {\it decouple} from those of the underlying
field. As we show here, the dynamical equations of the IGM within the
gravitational field of dark matter are the same as that used for
describing the statistical decoupling between the passive substance and
underlying field. Therefore, it is possible there is a statistical
discrepancy between the IGM field and underlying dark matter in
nonlinear regime.

The purpose of this paper is then two-fold. First we study whether
the intermittency of the QSO Ly$\alpha$ transmitted flux is matched by the
full hydro simulation samples. Secondly we search for possible
signs of the statistical discrepancy between the IGM field and
the underlying dark matter with these samples. The paper is organized as
follows. \S 2 presents a brief introduction to intermittent fields and the
method to quantify intermittent random fields. \S 3 describes the observed
samples that we use. \S 4 describes full hydrodynamic simulations. The
statistical deviation of the IGM distribution from
the dark matter field is shown in \S 5. A comparison between the model
predicted and observed intermittency
is performed in \S 6. Finally, the
conclusions and discussions are presented in \S 7. The dynamical equations
for the IGM evolution in non-linear regime is given in Appendix.

\section{Intermittency}

Much of this section has been presented in our previous work
(Pando et al 2002). For the paper to be self-contained, we repeat these
results briefly.
\subsection{The intermittent exponent}

Intermittency is used to characterize a random field with
`spike-gap-spike' features, i.e., structures that are essentially strong
enhancements randomly and widely scattered in
space and/or time, with a low field value between the spikes.
The spike-gap-spike feature is more pronounced on smaller scales.
This feature was originally found in the temperature and velocity
distributions in turbulence (Batchelor \& Townsend 1949; Frisch 1995).

An intermittent random density field $\rho({\bf x})$ is
defined by the asymptotic behavior (indicated below by $\asymp$) of the
ratio between the high- and low-order moments of the field as
(G\"artner \& Molchanov, 1990; Zel'dovich, Ruzmaikin, \& Sokoloff, 1990)
\begin{equation}
\frac{\langle [\rho({\bf x+ r})- \rho({\bf x})]^{2n}\rangle}
{[\langle [\rho({\bf x + r})- \rho({\bf x})]^2\rangle]^n}
\asymp\left \langle \frac{r}{L} \right \rangle^{\zeta},
\end{equation}
where $L$ is the size of the sample. Intermittency is measured by
the $n$- and $r$-dependencies of $\zeta$ -- the intermittent exponent.
If the exponent $\zeta$ is negative on small scales $r$, the field
is intermittent. That is, the ratio in eq. (1) for an intermittent
field diverges as $r\rightarrow  0$.

This divergence cannot be measured by individual higher
order correlation functions. Although three- and four-point
correlation functions are useful for distinguishing a
Gaussian from a non-Gaussian field, they are insensitive to the
difference between an intermittent field and a non-Gaussian, but
non-intermittent field. To measure intermittency, we define the
structure function as
\begin{equation}
S^{2n}_r = \langle|\Delta_r(x)|^{2n} \rangle
\end{equation}
where $\Delta_r(x)\equiv \rho(x+r)-\rho(x)$, and
$\langle...  \rangle$ is the average over the ensemble of
fields. If the field is statistically homogeneous, $S^{2n}_r$ is
independent of $x$ and depends only on $r$. $S^{2n}_r$ is called
the structure function. When $n=1$, we have
\begin{equation}
S^2_r =\langle |\Delta_r(x)|^{2} \rangle.
\end{equation}
$S^2_r$ is the mean of the square of the density fluctuations at
wavenumber $k\simeq 2\pi/r$, and therefore, $S^2_r$ is
the power spectrum of the field. With the structure function,
eq.(1) can be rewritten as
\begin{equation}
\frac{S^{2n}_r}{[S^{2}_r]^n} \propto
   \left(\frac{r}{L} \right )^{\zeta}.
\end{equation}
The ratio $S^{2n}_r/(S^{2}_r)^n$ is the n$^{th}$ moment $S^{2n}$
normalized by the power $S^2$. Since $S^{2}_r$ measures the width
of the probability distribution function (PDF) of $|\Delta_r(x)|$,
while $S^{2n}_r$ is sensitive to the tail of the PDF, the ratio
eq.(4) measures the fraction of density fluctuations in the tail on
different scales.

For an intermittent field the ratio of $S^{2n}_r$ to $[S^{2}_r]^n$
is larger for smaller $r$ and indicates that the field  contains
``abnormal'' events of large density fluctuations $|\Delta_r(x)|$. That
is, the PDF of $|\Delta_r(x)|$ is long-tailed. The long-tail events
correspond to a sharp increase or decrease of the density $\rho({\bf x})$.
They are rare events and yield high peaks or spikes in the distribution of
density difference. Generally speaking, the intermittent exponent $\zeta$
measures the smoothness of the field: for positive $\zeta$, the field is
smoother on smaller scales. If $\zeta$ is negative, the field is rough
on small scales.

The basic quantity used for calculating the intermittent exponent $\zeta$
is the density difference $|\Delta_r(x)|$. The exponent
$\zeta$ does not rely on the mean density $\bar{\rho}$ and can be
measured even if the mass field is fractal, in which case the
mean density is no longer available.

\subsection{Examples of the intermittent exponent}

In our previous work we presented several examples of intermittency in
various kinds of fields.  Here we summarize those results for the
kinds of fields that pertain to this work.

For a Gaussian field,
\begin{equation}
\frac{S^{2n}_r}{[S^{2}_r]^n} = (2n-1)!!.
\end{equation}
This ratio is independent of scale $r$, and therefore, the
intermittent exponent $\zeta=0$. A Gaussian field is not intermittent.
A self-similar field is also not intermittent.

For a hierarchically clustered field, it has been shown (Feng, Pando,
\& Fang 2001) that
\begin{equation}
 \frac{S^{2n}_r}{[S^{2}_r]^n} \propto \left(\frac{r}{L}
  \right )^{-(d-\kappa)(n-1)},
\end{equation}
where $d$ is the spatial dimension, and the coefficient $\kappa$ is a
constant depending on the power law index of the power spectrum. In the
case where $\kappa <d$, the field is intermittent with exponent
\begin{equation}
\zeta \simeq  -(d-\kappa)(n-1).
\end{equation}

Finally, for a lognormal field the PDF of $\Delta_r(x)$ is given by
\begin{equation}
P[\Delta_r(x)] =\frac{1}{2^{3/2}\pi^{1/2}|\Delta_r(x)|\sigma(r)}
 \exp\left \{ -\frac{1}{2}
\left (\frac {\ln|\Delta_r(x)|-\ln |\Delta_r(x)|_m}
{\sigma(r)}\right)^2  \right \},
\end{equation}
where $|\Delta_r(x)|_m$ is the {\it median} of $|\Delta_r(x)|$
(Vanmarcke 1983). The variance $\sigma(r)$ of $\ln|\Delta_r(x)|$
generally can be a function of the scale $r$.
Using eq.(8), we have
\begin{equation}
\frac{S^{2n}_r}{[S^{2}_r]^n} = e^{2(n^2-n)\sigma^2(r)}.
\end{equation}
The intermittent exponent of a lognormal field is then
\begin{equation}
\zeta \simeq 2(n^2-n)\sigma^2(r)/\ln(r/L).
\end{equation}
Because $r < L$, $\zeta$ is negative. Therefore, a lognormal field is
intermittent.

To summarize, the structure function and intermittent exponent provide
a complete and unified description of intermittent fields. The
$n$- and $r$-dependencies of the structure functions and intermittent
exponent $\zeta$ are sensitive to the details of the intermittency
of the field. These measures are very powerful for distinguishing among
fields that are Gaussian, self-similar, mono-fractal, multi-fractal, and
singular.

\subsection{The intermittent exponent in the wavelet basis}

The quantity $\Delta_r(x)$ or $[\rho(x+r)-\rho(x)]$ contains two
variables: the position $x$ and the scale $r$, and therefore,
$\rho(x+r)-\rho(x)$, can best be calculated with a space-scale
decomposition. Among the mathematical tools used to decompose
functions in scale space as well as physical space, the discrete
wavelet transform (DWT) is an excellent choice. The DWT
basis are orthogonal and complete, and therefore, the DWT
decomposition does not lose information (completeness), and does
not cause false correlations due to redundancy
(non-orthogonality). These two points are essential in studying
the higher order statistical features of random fields.

We restrict our discussion to a 1-D random field $\rho(x)$
extending in a physical or redshift spaces $L=x_2-x_1$. To apply the
DWT, we first chop the spatial range $L=x_2-x_1$
into $2^j$ subintervals labeled $l=0, ...2^j-1$ where $j$ is
a positive integer. Each subinterval spans a spatial range
$L/2^j$ with the $l^{th}$ subinterval spanning from $x_1+ Ll/2^j$ to
$x_1 + L(l+1)/2^j$. This operation decomposes the space $L$ into cells
$(j,l)$, where $j$ denotes the scale $L/2^j$, and $l$ the spatial
range.

Corresponding to each cell, there is a scaling function
$\phi_{jl}(x)$, and a wavelet function $\psi_{jl}(x)$. The scaling
function $\phi_{j,l}(x)$ is a window function on scale $j$ and at
position $l$. The wavelet functions $\psi_{jl}(x)$ form the basis for
the scale-space decomposition. The most important property of the
DWT basis is its locality in both scale and physical spaces.

With the DWT, a density field $\rho(x)$ can be decomposed as (Fang \&
Thews
1998).
\begin{equation}
\rho(x) = \sum_{l=0}^{2^j-1}\epsilon_{j,l}\phi_{jl}(x)+
 \sum_{j'=j}^{J} \sum_{l=0}^{2^{j'}-1}
  \tilde{\epsilon}_{j',l} \psi_{j',l}(x),
\end{equation}
where $J$ is given by the finest scale (resolution) of the sample,
i.e., $\Delta x=L/2^J$, and $j$ is the scale of interest. The scaling
function coefficient (SFC) $\epsilon_{jl}$ in eq.(11) is given
by projecting $\rho(x)$ onto $\phi_{j,l}(x)$
\begin{equation}
\epsilon_{j,l}=\int \rho(x)\phi_{j,l}(x)dx.
\end{equation}
The SFC $\epsilon_{jl}$ describes the mean field of the mode $(j,l)$.

The wavelet function coefficient (WFC), $\tilde{\epsilon}_{j,l}$,
in eq.(11) is obtained by projecting $\rho(x)$ onto $\psi_{j,l}(x)$
\begin{equation}
\tilde{\epsilon}_{j,l}= \int \rho(x) \psi_{j,l}(x)dx.
\end{equation}
The wavelet function $\psi_{jl}(x)$ is admissible (Daubechies 1992),
i.e., $\int \psi_{jl}(x)dx=0$, so the WFC is basically the difference
between $\rho(x)$ and $\rho(x+r)$, where
$x \simeq x_1+ lL/2^j$ and $r\simeq L/2^j$. Therefore, the WFC eq.(13)
measures the fluctuation on scale $j$ and at position $l$. Because
the admissibility of wavelets, eq.(13) can also be written as
\begin{equation}
\tilde{\epsilon}_{j,l}= \int \delta(x) \psi_{j,l}(x)dx,
\end{equation}
if the normalization is $\overline{\rho}=1$.  Thus, the WFC can
be used as the variable $\rho(x+r)-\rho(x)$ or $\delta(x+r)-\delta(x)$
in eq.(2). All compactly supported wavelet bases produce similar
results. We will use Daubechies 4 (Daubechies, 1992) in the study below.

When the ``fair sample hypothesis" is applicable (Peebles 1980), the
structure function eq.(2) can be calculated as the spatial average
\begin{equation}
S^{2n}_r =\frac{1}{L}\int |\Delta_r(x)|^{2n}dx,
\end{equation}
where $L$ is the spatial range of the sample. In the DWT representation,
eq.(15) yields
\begin{equation}
S^{2n}_j = \langle|\tilde{\epsilon}_{j,l}|^{2n} \rangle
=\frac{1}{2^j}\sum_{l=0}^{2^j-1}|\tilde{\epsilon}_{j,l}|^{2n},
\end{equation}
where $j$ plays the same role as $r$ in eq.(2). $S^{2n}_j$
is the mean of moment $|\tilde{\epsilon}_{j,l}|^{2n}$ over the
position index $l$. The intermittent exponent $\zeta$ in eq.(4)
can be calculated from
\begin{equation}
\frac{S^{2n}_j}{[S^{2}_j]^n} \propto  2^{-j\zeta}.
\end{equation}
Generally, $\zeta$ depends on $n$ and $j$.

\section{The Observed Ly$\alpha$ transmitted flux}

\subsection{QSO's Ly$\alpha$ absorption spectra}

The observed transmitted flux at wavelength $\lambda$ is
$F(\lambda)=F_c e^{-\tau(\lambda)}$, where $\tau(\lambda)$ is optical
depth,
and $F_c$ the continuum which is generally $\lambda$-dependent.
$F_c$ is given by the emission spectrum of the QSO considered.
To study the intermittent behavior of the IGM we focus on the fluctuation
field
$e^{-\tau(\lambda)}$. The intermittent features of transmitted flux of
QSO's Ly$\alpha$ absorption spectrum were previously studied with a set of
about
30 Keck HIRES spectra (Jamkhedkar, Zhan, \& Fang, 2000, Pando et al.
2002).
This work will consider the two best absorption spectra in that the sample
set.

\noindent{\it 1. HS1700+6416}

The absorption spectrum of QSO HS1700+6416 (redshift $z = 2.72$) covers a
wavelength range from 3727.012~\AA~to 5523.554~\AA, for a total of
55882 pixels. On average, a pixel is about 0.028~\AA. For all
pixels in this data set, the ratio
$\Delta \lambda/\lambda \simeq 7\times 10^{-6}$ and constant, or
in terms of the local velocity,
$\delta v =c\Delta \lambda/\lambda \simeq 2.1$ km/s, and therefore,
the resolution is about 4 km/s. The distance between $N$
pixels in units of the local velocity scale is given by
$\Delta v=2c(1-\exp{[-(1/2)N~\delta v/c]})$~km/s, or wavenumber
$k=2\pi /\Delta v$~s/km. As $\delta v /c \ll 1$, we have
$\Delta v \simeq N~\delta v$. In this work, we use the data from
$\lambda =$3815.6~\AA~ to 4434.3~\AA, which corresponds to
$z = 2.14 \sim 2.65$. The lower limit of the wavelength is set
to exclude Ly$\beta$ absorption, the upper to eliminate proximity
effects. The continuum $F_c$ for this spectrum was done
by IRAF CONTINUUM fitting by Kirkman and Tytler (1997).

In each pixel the data contain wavelength $\lambda_i$, flux
$F(\lambda_i)$ and noise $\sigma(\lambda_i)$, which accounts
for the Poisson fluctuations in the photon count, the noise due to
the background and the instrumentation. The S/N ratio is about 8.

This data set contains a fitted flux which smooths out all fluctuations
on smallest scales, i.e. one or two pixels. The fluctuations on these scales
are substantially suppressed by the fitting procedure. However, we analyze
only scales greater than or equal to 16 pixels. Therefore, the smoothing
has little effect on our results. Metal lines for this spectrum have been
identified down to the Doppler parameters $b \sim$ 10 km/s by Wei Zheng
(unpublished).

\noindent{\it 2. q0014+8118}

This spectra has the highest S/N$\sim $ 30 among the 30 Keck QSO spectra
collected by Kirkman and  Tytler (1997). The redshift of q0014+8118
is 3.387. For all pixels in this data set $\Delta \lambda/\lambda
\simeq 13.8 \times 10^{-6}$,  or $\delta v = c\Delta
\lambda/\lambda\simeq 4.01$~km/s, and, therefore, the resolution
is about 8~km/s. In each pixel, the data also contains wavelength
$\lambda_i$, flux $F(\lambda_i)$ and noise $\sigma(\lambda_i)$.
The continuum of the spectrum is given by IRAF CONTINUUM fitting
as well.

We use the wavelength region from the Ly$\beta$ emission to the
Ly$\alpha$ emission, excluding a region of about 0.06 in redshift close
to the quasar to avoid any proximity effects. This corresponds to
$z=2.702$ to
3.314, which contains 12703 pixels. The distance between $N$ pixels in
the units of
the local velocity scale is given by $\Delta v \simeq N\times 4.01$~km/s,
or wavenumber $k=2\pi /\Delta v$~s/km. Absorption features with $b <20$
km s$^{-1}$ of this spectrum have been identified.

\subsection{Treatment of unwanted data}

To study the intermittency of the transmission flux fluctuations, we
should
remove the effects of unwanted data, such as 1.) bad pixels (gaps
without
data), 2.) metal lines, and 3.) negative flux pixels. The last category
generally consists of saturated absorption regions having lower $S/N$. Although
the percentage of low $S/N$ data is not large, it will introduce large
uncertainties in the analysis.

The conventional technique for removing unwanted data is to eliminate
the unwanted bins and to smoothly rejoin the rest of the forest spectra.
However, by taking advantage of the localization of wavelets we can use
the algorithm of DWT de-noising by thresholding (Donoho 1995) as follows
\begin{enumerate}
\item Calculate the SFCs for both the transmission $F(\lambda)$ and
noise $\sigma(\lambda)$, i.e.
\begin{equation}
\epsilon^F_{jl}=\int F(x)\phi_{jl}(x)dx, \hspace{3mm}
\epsilon^N_{jl}=\int \sigma(x)\phi_{jl}(x)dx.
\end{equation}
\item Identify unwanted mode $(j,l)$ using the  condition
\begin{equation}
\left| \frac{\epsilon^F_{jl}} {\epsilon^N_{jl}} \right| <  f
\end{equation}
where $f$ is a constant. This condition flags all modes with
S/N less than $f$. We can also flag modes dominated by metal lines.
\item Since all the statistical quantities in the DWT representation
are based on an average over the modes $(j,l)$, we do not
count all the flagged modes when computing these averages.
\end{enumerate}
This is the conditional-counting method to treat
the unwanted data (Pando et al. 2002).

It should be emphasized that the condition (19) is applied at each
scale $j$. For instance, if the size of a saturated absorption region is $d$,
eq.(19) requires the removal of all modes $(j,l)$ on scales less than
$d$ and located within this region, as these modes have very low S/N
($\epsilon^F_{jl}/\epsilon^N_{jl} <  f$) and the local fluctuations in
these regions are not reliable. However, condition (19) does not
require the removal of the saturated absorption region when $d$ is
less than the scale of modes $(j,l)$ considered. If
$\epsilon^F_{jl}/\epsilon^N_{jl} > f$, the mode $(j,l)$ is counted in
the statistics, regardless of whether the mode $(j,l)$
contains the saturated absorption regions. The DWT analysis allows an
effective application of data containing low S/N regions. That is, with the
scale-by-scale denoising, it eliminates low S/N modes on small
scales and in these regions, but keeps their contributions to large scale
modes. With this procedure, no eliminating and smoothly rejoining of the
data is needed, and therefore, it avoids corrupting long wavelength modes.

With this method, we can still calculate the structure functions by
eq.(17), but the average is not over all modes $l$, but over the
un-flagged modes only. We can also flag modes $(j,l)$ one by one which
are dominated by metal lines. Since the DWT calculation assumes the sample
is periodized, this may cause uncertainty at the boundary. To reduce
this effect, we drop 3 modes near the boundary, and also flag two modes
around a unwanted mode to reduce any boundary effects of the chunks.

\subsection{Power spectrum of HS1700}

To check the quality of the data, we calculate the power spectrum
of the transmitted flux fluctuations,
$\Delta F=F(\lambda) - \langle F(\lambda)\rangle$, of HS1700+6416.
In the DWT representation, this power spectrum is given by (Pando
\& Fang 1998b; Fang \& Feng 2000, Jamkhedkar, Bi, \& Fang, 2001)
\begin{equation}
P_j=\frac{1}{2^j}\sum_{l=0}^{2^j-1}(\tilde{\epsilon}^F_{jl})^2
   - \frac{1}{2^j}\sum_{l=0}^{2^j-1}(\tilde{\epsilon}^n_{jl})^2.
\end{equation}
The first term on the r.h.s. of eq.(20) is the power of $\Delta F$
on scale $j$, in which the wavelet coefficients (WFC)
$\tilde{\epsilon}^F_{jl}$ are given by
\begin{equation}
\label{eq:WFC2}
\tilde{\epsilon}^F_{jl} = \int F(x)\psi_{jl}(x)dx.
\end{equation}
The second term on the r.h.s. of eq.(20) is due to the noise field
$\sigma(\lambda)$ and calculated by
\begin{equation}
(\tilde{\epsilon}^n_{jl})^2 = \int\sigma^2(x)\psi_{jl}^2(x)dx.
\end{equation}

The DWT power spectrum $P_j$ of a
random field $F(x)$ is related to its Fourier power spectrum $P(n)$  by
\begin{equation}
P_j = \frac{1}{2^j} \sum_{n = - \infty}^{\infty}
 |\hat{\psi}(n/2^j)|^2 P(n).
\end{equation}
where $\hat{\psi}$ is the Fourier transform of the wavelet.  This implies
that
the DWT power spectrum $P_j$ is the banded
Fourier power of the flux fluctuations $F(\lambda)-\langle
F(\lambda)\rangle$. Since for D4 wavelet $|\hat{\psi}(\eta)|^2$
is peaked at $\eta \simeq 1$, the band $j$ corresponds to the
wavenumber $k = 2 \pi n/L \simeq 2\pi 2^j/L$.

Figure 1 plots the results of the power spectra for HS1700+6416, in
which the data is treated three ways, 1.) using the original fitted flux,
2.) using the conditional counting method eq.(19) with $f=1$, and 3.) with
$f=3$.

At the first glance, the conditional-counting condition eq.(19) would seem
to preferentially miss--count modes in the low transmission regions and
lead to an $f$-dependence in the power spectrum. However Fig. 1 shows
that the power spectrum $P_j$ is independent of $f$ on entire the scale
range considered from $f=1$ to 3. The reason can be seen from
eq.(20), which shows that the contribution to the power $P_j$ given by
mode $(j,l)$ is $(\tilde{\epsilon}^F_{jl})^2-(\tilde{\epsilon}^n_{jl})^2$.
The noise subtraction term $(\tilde{\epsilon}^n_{jl})^2$ guarantees
that the contribution of modes with small ratio $S/N$ to $P_j$ is always
small or negligible. In other words, all the miss--counted modes
make very small or negligible contributions to $P_j$ regardless
the parameter $f$.

Also in Fig. 1, on scales $ j > 10$, or $\Delta v < 32 $ km/s, the power
of the fitted flux is lower than the power spectrum. This
discrepancy is probably because the fitted flux is over-suppressed,
i.e., not only noise, but also real fluctuations are suppressed by
the fitting procedure. This discrepancy indicates that the effects of
noise become significant on scales
$\Delta v < 32 $ km/s. For this reason, we use only data on scales
$\Delta v \geq 32 $ km/s. Sample q0014+8118 is of similar quality and
for it we also only use scales $\Delta v \geq 32 $ km/s.

\section{Hydrodynamical simulations}

\subsection{The simulation}

For the hydrodynamic simulations, we use a
Eulerian code based on a weighted essentially non-oscillatory (WENO)
scheme for a hyperbolic system (Jiang \& Shu, 1996; Shu, 1998).
WENO realizes the idea of adaptive stencils in the reconstruction
procedure based on the local smoothness of the numerical solution
to automatically achieve high order accuracy and non-oscillatory
property near discontinuities. It is extremely robust and stable
for solutions containing strong shock and complex solution
structures. This is especially important in studying
intermittency. In the context of cosmological applications, we
developed a hybrid N-body/hydrodynamical code that incorporates a
Lagrangian particle-mesh algorithm to evolve the collision-less
matter with the fifth order WENO scheme to solve the equation of
gas dynamics (Feng et.al, 2002).

In the present application, we computed the cosmic evolution of the
coupled system
of both dark matter and baryonic matter in a flat low
density CDM model ($\Lambda$CDM), which is specified by the
density parameter $\Omega_m=0.3$, cosmological constant
$\Omega_{\Lambda}=0.7$, Hubble constant $h=0.7$, and mass
fluctuation within a sphere of radius 8h$^{-1}$Mpc, $\sigma_8=0.9$. The
baryon fraction is fixed using the constraint from primordial
nucleosynthesis as $\Omega_b=0.0125h^{-2}$ (Walker et.al, 1991). The
linear power spectrum is taken from the fitting formulae presented by
Eisenstein \& Hu (1998). We perform the simulations in a 128$^3$ grid with
an equal number of dark matter particles. In order to examine the effects
of
numerical resolution, two simulations on a periodic, cubical box of size
6h$^{-1}$Mpc and 12h$^{-1}$Mpc are generated. The simulations start at a
redshift
$z=49$ and the results were outputed at the average redshifts of the each
observation samples (for q0014+8118, $z_m=2.915$, and HS1700+6416,
$z_m=2.329$). The timestep is chosen by the minimum value among three time
scales.
The first is from the Courant condition given by
\begin{equation}
 \delta t \le \frac{ cfl \times a(t) \Delta x}{\hbox{max}(|u_x|+c_s,
|u_y|+c_s, |u_z|+c_s)}
\end{equation}
where $\Delta x$ is the cell size, $c_s$ is the local sound speed, $u_x$,
$u_y$ and $u_z$
are the local fluid velocities and $cfl$ is the Courant number, typically,
we take $cfl=0.6$.
The second constraint is imposed by cosmic expansion which requires that
$\Delta a /a <0.02$
within single time step. The last constraint comes from the requirement
that a particle move
no more than a fixed fraction of cell size in one time step.

Atomic processes including ionization, radiative cooling and
heating are modeled as in Cen (1996) in a primeval plasma of hydrogen and
helium of composition ($X=0.76$, $Y=0.24$). The uniform UV-background of
ionizing photons is assumed to have a power-law spectrum of the form
$J(\nu)
=J_{21}\times
10^{-21}(\nu/\nu_{HI})^{-\alpha}$ergs$^{-1}$cm$^{-2}$sr$^{-1}
$Hz$^{-1}$, where the photoionizing flux is normalized by parameter
$J_{21}$ at the Lyman limit frequency $\nu_{HI}$, and is suddenly
switched on at $z\sim 6$ to heat the gas and reionize the universe.

One-dimensional fields are extracted along randomly selecting lines of sight
in the
simulation box. The density, temperature and velocity of the neutral gas
fraction on grids are firstly Gaussian smoothed using FFT techniques which
form the fundamental data set. The one-dimensional grid containing the
physical
quantities is further interpolated by a cubic spline. Using this
one-dimensional grid, the optical depth $\tau$ at each pixel is then
obtained by integrating in real space and we  include the effect of the
peculiar
velocity and convolve with Voigt thermal broadening. To have a fair
comparison with observed spectra, $\tau$ was Gaussian smoothed to match
with the spectral resolutions of observation. The transmitted flux
$F=\exp(-\tau)$ is normalized such that the mean flux decrement in the
spectra match with observations.

Each mock spectrum is sampled on a $2^{10}$ grid with the same spectral
resolution as the
observation. As the corresponding comoving scale for $2^{10}$ pixels is
larger than the simulation box size, we replicate periodically the sample. To
achieve the greatest the statistical
independence, we randomly change the direction of line of sight while
crossing the boundary of the simulation box.
For each observed spectrum, 1000 mock spectra are generated. In addition,
we also
output 100 samples of the density distribution of both dark matter and
neutral hydrogen on the one-dimensional grid.

\subsection{Power spectrum of hydrodynamical simulation samples}

Figure 2 compares the DWT power spectra measured in the mock samples and the
observed data of HS1700. The comoving wavelength
$k$ in Fig. 2 is given by $k = 2\pi/L$, where $L = (\Delta v
/H_0)[\Omega_m(1+z_m)^3+\Omega_{\Lambda}]^{-1/2}$. The 1-$\sigma$ errors
for the simulation samples are derived from the 1000 mock spectra. For the
observed spectrum of HS1700, we divide the first $2^{14}$ pixels of the
observed spectrum into 16 subsections, each with $2^{10}$ pixels. The DWT
power spectra are obtained for each subsection respectively, and the
1-$\sigma$ error is estimated from the variance of the 16 sections. [For an
intermittent field, the error estimation of power spectrum is sensitively
dependent on how one treats the intermittency (Jamkhedkar, 2002). We do
not discuss the details of this point here as it will not affect the
following discussion.]

Figure 2 shows that simulation samples with the 12 Mpc box are
in good agreement with observations on all scales $k \leq 15$ h
Mpc$^{-1}$, which corresponds to $j=10$ in Fig. 1. The simulation
samples with the  6 h$^{-1}$Mpc box have higher power than
observed sample on small scales, but lower power on large scales.
This is expected because the simulation in the 6 h$^{-1}$Mpc box
lacks perturbations on scales $k \leq 1$ h Mpc$^{-1}$ and
the normalization of the mean flux decrement will lead to the
higher power on small scales to compensate for the loss of powers
on large scales.

On small scales $k >15$ h Mpc$^{-1}$, the simulated power is
significantly lower than the observation data. This shows again
that noise contamination becomes important on scales $k >15$ h
Mpc$^{-1}$, or $j > 10$. Therefore, as in \S 3.3, we use
the simulation data only on scales $k >15$ h Mpc$^{-1}$, or $j
\leq 10$, or $\Delta v \geq 32 $ km/s.

\section{Statistical discrepancy between baryonic gas and dark matter}

\subsection{Relation between IGM and dark matter mass fields}

It is well known that in the linear approximation the density
contrast of the IGM, $\delta_b({\bf x}, t)$ is equal to that of dark
matter
$\delta_{DM}({\bf x}, t)$ on scales larger than the Jeans length. This is
true even if initially $\delta_b({\bf x}, t_0) \neq \delta_{DM}({\bf x},
t_0)$
(Bi, B\"orner \& Chu 1993, Nusser 2000, Matarrese \& Mohayee 2002).
In other words, the density fluctuations of the IGM trace the dark matter
fields point-by-point. Since the clustering of the IGM is governed by the
gravity of dark matter, the IGM can be considered as "passive substance"
with
respect to the underlying dark matter. Thus, the point-by-point relations
$\rho_b({\bf x}, t) \propto \rho_{DM}({\bf x}, t)$ and
$\delta_b({\bf x}, t) = \delta_{DM}({\bf x}, t)$ on scales larger than
the Jeans length also looks reasonable in the nonlinear regime. In this
case, {\it all} the statistical properties of the IGM field on
scales larger than the Jeans length are the same as the underlying dark
matter field.

However, it is known that a passive substance in a hydrodynamic
system does not always behave like the underlying dominant mass field
when non-linear evolution takes place. The statistical properties
of a passive substance can {\it decouple} from those of the
underlying field in the nonlinear regime. Moreover, the evolution of
the IGM driven by the gravitational field of dark matter is described
by a non-linear equation which is essentially the same as that used for
modeling the statistical decoupling between a passive substance and its
underlying field (see Appendix A.) Hence, we expect to see a statistical
discrepancy between $\delta_b({\bf x}, t)$ and $\delta_{DM}({\bf x}, t)$
on scales larger than the Jeans length.

To determine whether any statistical decoupling has occurred, we begin by
assuming that $\delta_b({\bf x}, t) = \delta_{DM}({\bf x}, t)$. Then wavelet
transforming yields $\tilde{\epsilon}^{b}_{jl}=\tilde{\epsilon}^{DM}_{jl}$,
and therefore
\begin{equation}
\langle \tilde{\epsilon}^{b}_{jl}\tilde{\epsilon}^{DM}_{jl}\rangle
  = \langle [\tilde{\epsilon}^{b}_{jl}]^2 \rangle =
   \langle[\tilde{\epsilon}^{DM}_{jl}]^2\rangle.
\end{equation}
$\langle [\tilde{\epsilon}^{b}_{jl}]^2 \rangle$ and
$\langle[\tilde{\epsilon}^{DM}_{jl}]^2\rangle$ are, respectively,
the DWT power spectrum of the IGM and dark matter fields. Therefore,
the first conclusion of the point-by-point relation
$\delta_b({\bf x}, t) = \delta_{DM}({\bf x}, t)$ is that the power spectra
of the IGM and dark matter should be the same. Figure 3 indeed shows that
the power spectra of the IGM and dark matter fields are about the same.

With eq.(24), it is obvious that the power spectrum of the field
$\delta_b({\bf x}, t) - \delta_{DM}({\bf x}, t)$
should be equal to zero, or considering noise, very small.
Figure 4 gives the power spectra
$\langle[\tilde{\epsilon}^{b}_{jl}- \tilde{\epsilon}^{DM}_{jl}]^2\rangle$
and $\langle[\tilde{\epsilon}^{b}_{jl}]^2\rangle$ and shows that
on large scales $j< 8$ ($\Delta v > 128$ km/s)
$\langle[\tilde{\epsilon}^{b}_{jl}- \tilde{\epsilon}^{DM}_{jl}]^2\rangle$
is much less than $\langle[\tilde{\epsilon}^{b}_{jl}]^2\rangle$. The ratio
of $\langle[\tilde{\epsilon}^{b}_{jl}-
\tilde{\epsilon}^{DM}_{jl}]^2\rangle$
to $\langle[\tilde{\epsilon}^{b}_{jl}]^2\rangle$ is $< 0.05$, which
probably is due to noise. However,
on scales $j\geq 8$ ($\Delta v \geq 128$ km/s), this ratio gradually
increases to $\simeq 0.4$. This difference between large and small scales
cannot be explained by the noise, as can be seen from Fig. 2 in which the
error bars in the range $1 < k <10$ h$^{-1}$
Mpc (corresponding to $j=7$ to 10) are about the same. Figure 4 indicates
that $\langle \tilde{\epsilon}^{b}_{jl}\tilde{\epsilon}^{DM}_{jl}\rangle
 < \langle [\tilde{\epsilon}^{b}_{jl}]^2 \rangle$ on small scales, i.e.,
the point-by-point correlation between the IGM density fields and dark
matter fields is weaker on small scales. This is a sign of the
statistical discrepancy between the IGM and dark matter mass fields.

We can further make the case by calculating the structure functions for
the field $\delta_b({\bf x}, t) - \delta_{DM}({\bf x}, t)$. The result is
plotted in Fig. 5 and shows that for $j=6$ the ratio of the
moments is Gaussian, i.e. $\ln S^{2n}_j/(S^{2}_j)^n = \ln (2n-1)!!$
[eq.(5)].  On scales $j\geq 8$, $\ln S^{2n}_j/(S^{2}_j)^n$ is significantly
larger than a Gaussian distributions which strongly indicates of the
statistical discrepancy between $\delta_b({\bf x}, t)$ and
$\delta_{DM}({\bf x},t)$ is not due to noise. For a given $n$,
$\ln S^{2n}_j/(S^{2}_j)^n$ increases with scale $j$ so that the
intermittent exponent $\zeta$ [eq.(21)] is non-zero and negative, and
therefore, the field $\delta_b({\bf x}, t) - \delta_{DM}({\bf x}, t)$ is
intermittent. This result is consistent with the dynamical equation (A14),
which requires that the difference
$\delta_b({\bf x}, t) - \delta_{DM}({\bf x}, t)$ generally is an
intermittent field. Thus, we conclude again that there is statistical
discrepancy from the point-by-point relation
$\delta_b({\bf x}, t) = \delta_{DM}({\bf x}, t)$.

\subsection{Relation between HI and dark matter}

The statistical discrepancy in the relation between the density fields
of neutral hydrogen and dark matter is important as well. For instance,
the pseudo-hydro technique assumes that
$\delta_{HI}({\bf x}, t) \propto \delta_{DM}^{a}({\bf x}, t)$
with $1.5 < a < 1.9$ (Hui \& Gnedin, 1997). This relation yields
$\langle \tilde{\epsilon}^{HI}_{jl}\tilde{\epsilon}^{DM}_{jl}\rangle >0$,
and therefore the power spectrum
$\langle[\tilde{\epsilon}^{HI}_{jl}+ \tilde{\epsilon}^{DM}_{jl}]^2\rangle$
should be larger than
$\langle[\tilde{\epsilon}^{HI}_{jl}-
\tilde{\epsilon}^{DM}_{jl}]^2\rangle$.
Figure 6 presents the power spectra
$\langle [\tilde{\epsilon}^{HI}_{jl}\pm
\tilde{\epsilon}^{DM}_{jl}]^2\rangle$
and shows
$\langle[\tilde{\epsilon}^{HI}_{jl}+ \tilde{\epsilon}^{DM}_{jl}]^2\rangle
>
\langle[\tilde{\epsilon}^{HI}_{jl}- \tilde{\epsilon}^{DM}_{jl}]^2\rangle $
on large scales, but this difference is significantly smaller on smaller
scales.

Figure 7 gives the power spectra
$\langle[\tilde{\epsilon}^{HI}_{jl}- \tilde{\epsilon}^{DM}_{jl}]^2\rangle
$
and $\langle [\tilde{\epsilon}^{b}_{jl}]^2 \rangle$ and reveals that it is
similar to Fig. 4. All the results indicate that the positive correlation
$\langle \tilde{\epsilon}^{HI}_{jl}\tilde{\epsilon}^{DM}_{jl}\rangle >0$
becomes weaker on scales $j \geq 8$ or $k < 3$ h Mpc$^{-1}$ or
$L > 2$ h$^{-1}$ Mpc. Since intermittent features become significant on
scales $j \geq 8$, the dynamical assumption that
$\rho_{HI}({\bf x})\propto \rho_{DM}^{a}({\bf x})$ is applicable
only for lower order statistics (like the power spectrum.)

\section{Comparison of intermittent features between real and
  simulation samples}

\subsection{Structure functions of real and simulation samples}

We now calculate the structure functions for the
transmitted flux fluctuations of the simulation samples. The
results are illustrated in Fig. 8 for HS1700 and Fig. 9 for q0014. The
error bars are the 1-$\sigma$ deviations estimated from the 1000
simulation samples. Comparing Figs. 8 and 9 with Fig. 2, we see
that $\ln S^{2n}_j/(S^{2}_j)^4$ with
$n=2$ has smaller errors than the power spectrum. Even
at 6th and 8th orders, the $j$-dependence of $\ln S^{2n}_j/(S^{2}_j)^n$
can be very well fitted by a line. This implies that to measure the higher
order statistical behavior of an intermittent field, the structure
functions are effective and stable.
For an intermittent field, the large uncertainty in the power spectrum
are caused by rare and improbable high spikes in the
fluctuations. However, the structure function is an ensemble
average of the ratio between $S^{2n}_j$ and $(S^{2}_j)^n$ and this
reduces the effect of individual high spikes.

Figures 8 and 9 also show that for a given $n$, $\ln
S^{2n}_j/(S^{2}_j)^n$ increases with scale $j$ so that  the
intermittent exponent $\zeta$ [eq.(21)] is non-zero and negative.
The transmitted flux fluctuations are, hence,
intermittent. In Fig. 10, we show $\ln S^{2n}_j/(S^{2}_j)^n$ vs.
$j$ of HS1700 mock samples produced by both the  12 and 6 h$^{-1}$Mpc box.
The 12 h$^{-1}$Mpc samples is a little higher $\ln
S^{2n}_j/(S^{2}_j)^n$, but there are no significant differences between
the simulation samples on scales considered.

The structure functions measured for the real data of HS1700 is also
shown in Fig. 10, in which the error bars  HS1700 are found by
bootstrap re-sampling. Within the error bars, the
observed structure function is consistent with the 12
Mpc box simulation samples for the LCDM model. Although the
consistency shown in Fig. 10 is still not
perfect, it is substantially better than the results obtained from
the pseudo hydro simulations. Pando et al (2002)
found that the structure functions of the pseudo hydro
simulation samples were generally larger than those of the real data
[see Fig. 7 of Pando et al.  (2002)].

The structure functions for both HS1700 and q0014 have similar
behavior although they lie at different redshifts. The
result demonstrates that one cannot see a significant redshift
evolution in the range of $2 < z <3$ with the $j$-dependence of
the structure functions.

\subsection{$n$-dependence of the structure function}

We now turn to the $n$-dependence of the structure functions.
Figures 11 and 12 are, respectively, $\ln S^{2n}_j/(S^{2}_j)^n$ vs. $n$
for the simulation samples and real data of HS1700 and q0014 on scales
$j=9$ and 10. The figures show again that the simulation result is
consistent with observations on the scales considered.

For a Gaussian field, the $n$-dependence of structure function is
$\ln S^{2n}_j/(S^{2}_j)^n =\ln (2n-1)!!$ [eq.(5)], which is also
plotted in Figs. 11 and 12.  We note, from Figs. 11 and 12, that although
$\ln S^{2n}_j/(S^{2}_j)^n$ is above that of
a Gaussian field,  $\ln S^{2n}_j/(S^{2}_j)^n$
seems to ``parallel" the Gaussian curve. This means that
although the field on scale $j=9$ is non-Gaussian, the PDF of
$\tilde{\epsilon}^F_{jl}$ doesn't have a significantly long tail. On
scale $j=10$, the slope of $S^{2n}_j/(S^{2}_j)^n$ becomes
larger than $(2n-1)!!$, which indicates that the PDF of
$\tilde{\epsilon}^F_{jl}$ tends to be long tailed.

The $n$-dependence of  $\ln S^{2n}_j/(S^{2}_j)^n$ appear to be more
sensitive to redshift evolution. Figure 11 shows that the
relation of $S^{2n}_j/(S^{2}_j)^n$ for $j=9$ is clearly
above the Gaussian field, while for Fig. 12 it is only marginally
different from a Gaussian field. Thus, the intermittent behavior
on scale $j=9$ probably developed at redshift $z \leq 3$.

Similar to the power spectrum (Fig. 1), the results of Figs. 8 - 12
result is $f$-independent on the entire scale range considered
for $f=1$ to 3 (Pando et al 2002). To see the
effect of metal lines, we plot Fig. 13, which shows the $n$-dependence
of $\ln S^{2n}_j/(S^{2}_j)^n$ ($j=10$) with and without metal line
removal. The result for $j=10$ is less affected by the metal line
removing because most metal lines have width less than
$\Delta v = 32$ km/s ($j=10$).

\subsection{Intermittent exponent}

The $n$-dependence of the intermittent exponent $\zeta$ can be calculated
by eq.(17), which gives
\begin{equation}
\zeta(n) = -\frac{1}{j}\ln_2 \frac{S^{2n}_j}{(S^{2}_j)^n} + {\rm const},
\end{equation}
or
\begin{equation}
\zeta(n) - \zeta(1) = -\frac{1}{j}\ln_2 \frac{S^{2n}_j}{(S^{2}_j)^n}.
\end{equation}
For the simulation samples, the relation of
$\ln_2 S^{2n}_j/(S^{2}_j)^n$ vs. $n$ shown in Fig. 10 can
be very well fitted by a line in the range from  $j=7$ to 10 (or
$2 \leq k \leq 16$ h Mpc$^{-1}$), and therefore,
one can find a constant $\zeta(n)$ for each given $n$. The real data
is not as well fitted over the same scales, but we nonetheless show
the results in figures 14 and 15.

Again the full hydro simulation shows better result than the pseudo hydro
simulation. The value of $|\zeta|$ for the real
data is found to be systematically lower than that given by pseudo
hydro simulation for the LCDM models (Pando at al. 2002). Although, the
value $|\zeta|$ from the hydro simulation is still higher than real data,
the differences are no longer larger than the error bars for all
$n$. Figure 16 shows the intermittent exponent for samples with 12 and 6
h$^{-1}$Mpc boxes. Note that the conclusions drawn from Figs. 14 and
15 hold also for samples with the 6 h$^{-1}$Mpc box.

\subsection{Intermittency of HI density field}

It has been argued that the intermittency of the Ly$\alpha$
transmitted flux field $F(\lambda)$ may not indicate intermittency
in the IGM distribution $\delta_{b}(\lambda)$ or HI distribution
$\delta_{HI}(\lambda)$. The reason is that in the exponential relation
between the flux and optical depth $F(\lambda)=F_ce^{-\tau(\lambda)}$,
and the optical depth $\tau(\lambda)$ is approximately related to
$\delta_{b}(\lambda)$ by
\begin{equation}
\tau(\lambda) =
    A[\rho_b(\lambda)/\bar{\rho}_b]^a=A[1+\delta_b(\lambda)]^a,
\end{equation}
where the parameter $A$ depends on the cosmic baryonic density, the
photoionization rate of HI, and the temperature of IGM. Therefore, even
when $\delta_b(\lambda)$ or $\delta_{HI}(\lambda)$ are Gaussian, the
PDF of the flux $F(\lambda)$ might be lognormal due to the exponential
relation between
$F(\lambda)$ and $\delta_{HI}(\lambda)$.

However as has been emphasized in \S 2, the intermittency of
$F(\lambda)$ field is not measured by the PDF of $F(\lambda)$, but
its difference $\Delta_{\delta \lambda} F(\lambda) \equiv
F(\lambda +\delta \lambda)-F(\lambda)$. The structure functions
are defined by the density difference. Thus, we have
approximately
\begin{equation}
\Delta_{\delta \lambda} F(\lambda) \simeq - e^{-\tau(\lambda)}
 \Delta_{\delta \lambda}\tau(\lambda)
 \end{equation}
where $\Delta_{\delta \lambda}\tau(\lambda)\equiv \tau(\lambda
+\delta \lambda)-\tau(\lambda)$. Using eq.(27), we have
\begin{equation}
\Delta_{\delta \lambda}\tau(\lambda) \simeq -A
a[1+\delta_H(\lambda)]^{a-1}
\Delta_{\delta \lambda}\delta_b(\lambda),
\end{equation}
where $\Delta_{\delta \lambda}\delta_b(\lambda) \equiv
\delta_b(\lambda+\delta \lambda) - \delta_b(\lambda)$.

Eqs.(29) and (30) show that if the transmitted flux fluctuations are
intermittent, the baryonic matter density field and HI field
should also be intermittent (Zhan, Jamkhedkar, \& Fang 2001).
Figure 17 presents the structure functions for the HI field of the 12 Mpc
box samples simulation and shows that
the HI field is indeed intermittent. In fact, this field has larger structure
functions than the flux fluctuations probably because the transmitted flux
cannot trace the intermittency of optical depth at heavy absorption regions.

\section{Discussion and conclusions}

We analyzed the intermittent behavior of the Ly$\alpha$ transmitted flux
fluctuations of real data and samples produced by full hydro simulations.
This analysis covers scales from $\sim$ 10 h$^{-1}$ Mpc to 400
h$^{-1}$ kpc ($j= 6$ to 10), and statistical orders $2n$ from 2 to 8. The
intermittent behavior is found to be significant on scales less than about
$\sim$ 1 h$^{-1}$ Mpc, and redshift $z\leq 3$. Over the entire scale range and
statistical orders considered, the intermittent behavior of the real data
matches the samples produced by full hydrodynamic simulations in the
LCDM model.

We have shown that the point-by-point correlation of the IGM and HI fields
with dark matter field becomes weaker on scales less than about 2 h$^{-1}$ Mpc,
which is much larger than the Jeans length at the mean density. Although
the power spectrum of the IGM field is still about the same as that of dark
matter at these scales, the mass density field of the IGM is no longer
point-by-point  proportional to the underlying dark matter field.

By using intermittency we have detected a the dynamical discrepancy
between the IGM field and the dark matter distribution. The study of
cosmic large scale structure has mainly concentrated on either the
two-point correlation function and its scaling, or on the massive collapsed
halos of dark matter. Intermittency may bridge the gap between the two
approaches and provide new, physically interesting insights in cosmic
clustering.

\acknowledgments

We thank Dr. D. Tytler for kindly providing the data of the Keck
spectrum HS1700+64 and q0014+8118, and Dr. W. Zheng for
providing the metal-line identification of HS1700+64. Dr. Jamkhedkar is
acknowledged for her help, especially, Figure 1 which is taken from her
thesis. We thank also Dr. D. Weinberg for comments and suggestions in
his referee's report. LLF acknowledges support from the National
Science Foundation of China (NSFC) and National Key Basic Research
Science Foundation.

\appendix

\section{Dynamical equations of the IGM and dark matter mass field}

\subsection{Basic equations}

Let us consider a flat universe having cosmic factor $a(t)\propto t^{2/3}$
and dominated by dark matter. We describe the dark matter by a mass
density
field $\rho({\bf x}, t)$ and a peculiar velocity field ${\bf v}({\bf x},
t)$,
where ${\bf x}$ is the comoving coordinate. In hydrodynamical
descriptions,
the equations of dark matter consist of the continuity, the momentum,
and the gravitational potential equations as follows
(Wasserman 1978)
\begin{equation}
\frac{\partial \delta}{\partial t} +
  \frac{1}{a}\nabla \cdot (1+\delta) {\bf v}=0
\end{equation}
\begin{equation}
\frac{\partial a{\bf v}}{\partial t}+
 ({\bf v}\cdot \nabla){\bf v}= -\nabla \phi
\end{equation}
\begin{equation}
\nabla^2 \phi = 4\pi G a^2\bar{\rho}\delta
\end{equation}
where density perturbation
$\delta({\bf x},t)=[\rho({\bf x}, t) - \bar{\rho}]/\bar{\rho}$, and the
mean density $\bar {\rho} =1/6\pi Gt^2 \propto a^{-3}$. The peculiar
gravitational potential $\phi$ is zero (or constant) when density
perturbation $\delta=0$. The operator $\nabla$ is acting on the comoving
coordinate ${\bf x}$. For the growth mode of the perturbations, the velocity
is irrotational. In this case, one can define a velocity potential by
\begin{equation}
{\bf v}=- \frac {1}{a}\nabla \varphi
\end{equation}
With this potential, the momentum equation (A2) can be rewritten as
the Bernoulli equation
\begin{equation}
\frac{\partial \varphi}{\partial t}-
\frac{1}{2a^2}(\nabla \varphi)^2 = \phi.
\end{equation}

Since there is only gravitational interaction between dark matter
and the IGM or cosmic baryonic gas, it is convenient to describe
the IGM by its mass density field $\rho_b({\bf x},t)$ and velocity
field ${\bf v}_b({\bf x},t)$. The hydrodynamical equations of the IGM
are
\begin{equation}
\frac{\partial \delta_b}{\partial t} +
  \frac{1}{a}\nabla \cdot (1+\delta_b) {\bf v}_b=0
\end{equation}
\begin{equation}
\frac{\partial a{\bf v}_b}{\partial t}+
 ({\bf v}_b\cdot \nabla){\bf v}_b=
-\frac{1}{\rho_b}\nabla p_b - \nabla \phi
\end{equation}
where the density perturbation of the IGM
$\delta_b({\bf x},t)= [\rho_b({\bf x},t)-\bar{\rho}_b]/\bar{\rho}_b$, and
$\bar{\rho}_b$ the mean density of the IGM. The gravity of the IGM is
negligible. The evolution of the IGM is governed by the gravitation of
dark matter, and therefore, the gravitational potential $\phi$ in eq.(A7)
is still given by eq. (A3).

To sketch the gravitational clustering of the IGM, we will not consider
the details of heating and cooling. Thermal processes are generally local,
and therefore, it is reasonable to describe the thermal processes
by a polytropic relation $p_b({\bf x},t) \propto \rho_b^{\gamma}({\bf
x},t)$.
Thus eq.(A7) becomes
\begin{equation}
\frac{\partial a{\bf v}_b}{\partial t}+
 ({\bf v}_b\cdot \nabla){\bf v}_b=
  -\frac{\gamma k_B T}{\mu m_p} \frac{\nabla \delta_b}{(1+\delta_b)}
  - \nabla \phi
\end{equation}
where the parameter $\mu$ is the mean molecular weight of IGM
particles, and $m_p$ the proton mass. In this case, we don't need the
energy
equation and the IGM temperature evolves as $T \propto \rho^{\gamma-1}$,
or $T =T_0(1+\delta_b)^{\gamma-1}$. Eq.(A8) is different from eq.(A2)
only by the term with temperature $T$. If we treat this term in the
linear approximation, we have
\begin{equation}
\frac{\partial \varphi_b}{\partial t}-
\frac{1}{2a^2}(\nabla \varphi_b)^2 - \frac{\nu_b}{a^2}\nabla^2 \varphi_b
=\phi,
\end{equation}
where $\varphi_b$ is the velocity potential for IGM field defined by
\begin{equation}
{\bf v}_b= - \frac {1}{a}\nabla \varphi_b.
\end{equation}
The coefficient $\nu_b$ is given by
\begin{equation}
\nu_b=\frac{\gamma k_BT_0}{\mu m_p (d \ln D(t)/dt)},
\end{equation}
where $D(t)$ describes the linear growth behavior.

Eq.(A9) shows that the clustering of the IGM field is completely controlled
by the gravity of underlying mass fields of dark matter. It can then be
considered as a "passive substance" with respect to the underlying dark
matter. The term with $\nu_b$ in eq.(A9) acts like a viscosity
which is due to thermal diffusion characterized by the Jeans length
$k_J^2=(a^2/t^2)(\nu m_p/\gamma k_BT_0)$. In linear regime, eqs.(A5) and
(A9) yield solution $\varphi_b = \varphi$ on scales larger than the Jeans
length.

\subsection{Equation of the difference between the IGM and dark matter
fields}

To study the possible difference between the IGM and dark matter, we
define
a variable $\Theta$ by
\begin{equation}
\Theta = \varphi_b -  \varphi
\end{equation}
which describes the deviation of IGM from dark matter. With eqs. (A5)
and (A9), we have
\begin{equation}
\frac{\partial \Theta}{\partial t}-
\frac{1}{2a^2}(\nabla \Theta)^2 - \frac{\nu_b}{a^2}\nabla^2 \Theta
 = \frac{\nu_b}{a^2}\nabla^2 \varphi +
\frac{1}{a^2}(\nabla \Theta)(\nabla \varphi).
\end{equation}
When $\nu_b=0$, this has solution $\Theta=$ const, if the initial
condition is $\Theta=$ const. During linear evolution, we have
$\Theta= 0$, and therefore $\Theta= 0$ is the initial condition
for eq.(A13). Thus, the solution $\Theta=0 $ is correct even in
the nonlinear regime if $\nu_b=0$. There is no deviation of IGM
from dark matter.

In the case of $\nu_b \neq 0$, $\Theta=0 $ will no longer be the solution
of eq.(A13) even if it is $\Theta=0$ initially. Considering that the
potential $\varphi$ of dark matter essentially is stochastic, the r.h.s.
of eq.(13) plays the role of stochastic forces. It drives $\Theta$ to be
non-zero. This leads to a statistical or stochastic discrepancy of the IGM
field from dark matter field. This discrepancy is not trivial even when we
consider the linear approximation of eq.(A13) with respect to $\Theta$,
i.e.
\begin{equation}
\frac{\partial \Theta}{\partial t}
 - \frac{1}{a^2}(\nabla \Theta)(\nabla \varphi)-
\frac{\nu_b}{a^2}\nabla^2 \Theta = \frac{\nu_b}{a^2}\nabla^2 \varphi.
\end{equation}
Eq.(14) essentially is the same as that widely used to dynamically
model the discrepancy, or decoupling, of the statistical properties of the
passive substance from the underlying field (for a review, Shraiman \&
Siggia, 2001). Since the advection term
$\frac{1}{a^2}(\nabla \Theta)(\nabla \varphi)$ doesn't depend on the
diffusion scales, the statistical discrepancy of the passive substance
from the underlying field can appear on scales larger than the
diffusion scale.

The field $\Theta$ given by eqs.(A13) and (A14) is generally intermittent.
It has been shown that the field $\Theta$ given by eq.(A14) will be
intermittent, even when the field $\varphi$ is Gaussian (Kraichnan, 1994).
Eq.(A13) is similar to the stochastic-force-driven Burgers' equation,
or the so-called KPZ equation (Kardar, Parisi, \& Zhang 1986, Berera \&
Fang 1994, Barab\'asi, \& Stanley, 1995, Jones 1999), which is typical of
the dynamical model of intermittency (Polyakov 1995; Balkovsky et al.
1997; E et al. 1997). Therefore, the difference between the IGM and dark
matter fields, $\Theta$, is probably intermittent.

\clearpage

\begin{figure}
\figurenum{1} \epsscale{0.8} \plotone{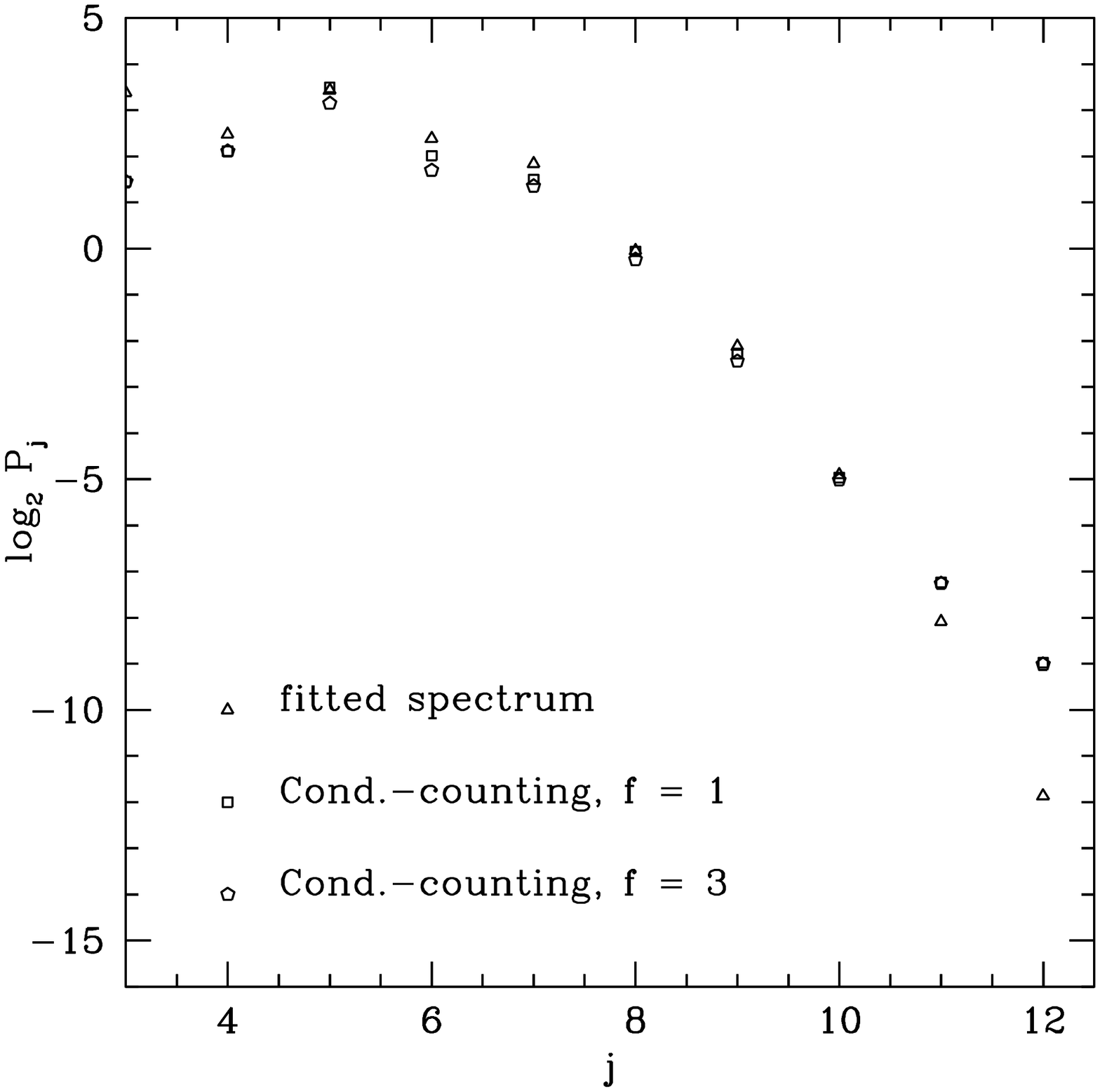} \caption{The power
spectra of the HS1700+6416 transmitted flux for 1.) the fitted
flux (triangle), 2.) conditional counting with $f=1$ (square), and
3.) with $f=3$ (pantegan). The physical scale for $j$ is $\Delta v
= 2^{13-j} \times 4$ km/s.}
\end{figure}

\begin{figure}
\figurenum{2} \epsscale{0.9} \plotone{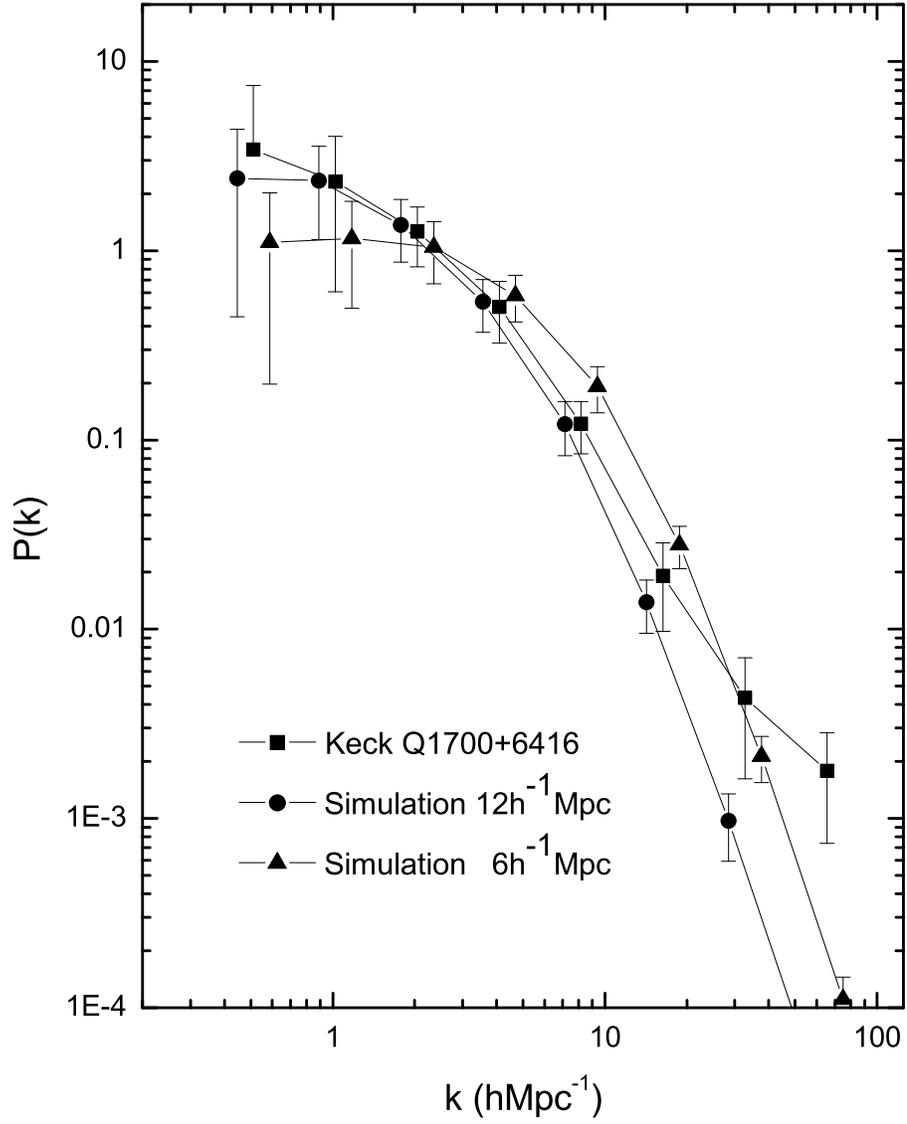}\vspace{-1.5cm} \caption{The power
spectra of HS1700+6416 for 1.) real data (square); 2.) hydro
simulation in a 12 Mpc box (circle), and 3.) in 6 Mpc box
(triangle). The comoving wavenumber $k$ is calculated in the LCDM
model. For clarity, the power spectra of the simulations are
shifted slightly.}
\end{figure}

\begin{figure}
\figurenum{3} \epsscale{0.8} \plotone{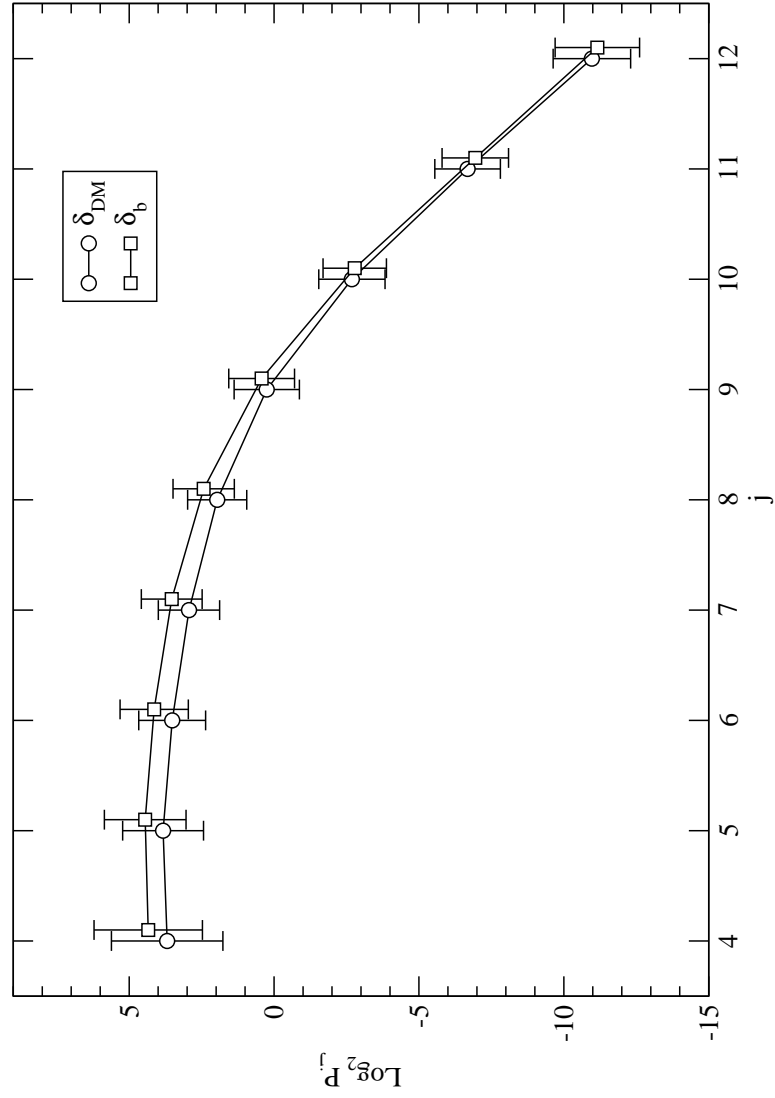}\caption{Power spectra
of IGM mass field $\delta_{b}$ (square) and dark matter
$\delta_{DM}$ (circle) from the hydro simulation in the 12 Mpc
box. The mean and variance are calculated from the 100
realizations. The physical scale for $j$ is $\Delta v = 2^{13-j}
\times 4$ km/s. For clarity, the power spectrum of the IGM is
shifted slightly. }
\end{figure}

\begin{figure}
\figurenum{4} \epsscale{0.8} \plotone{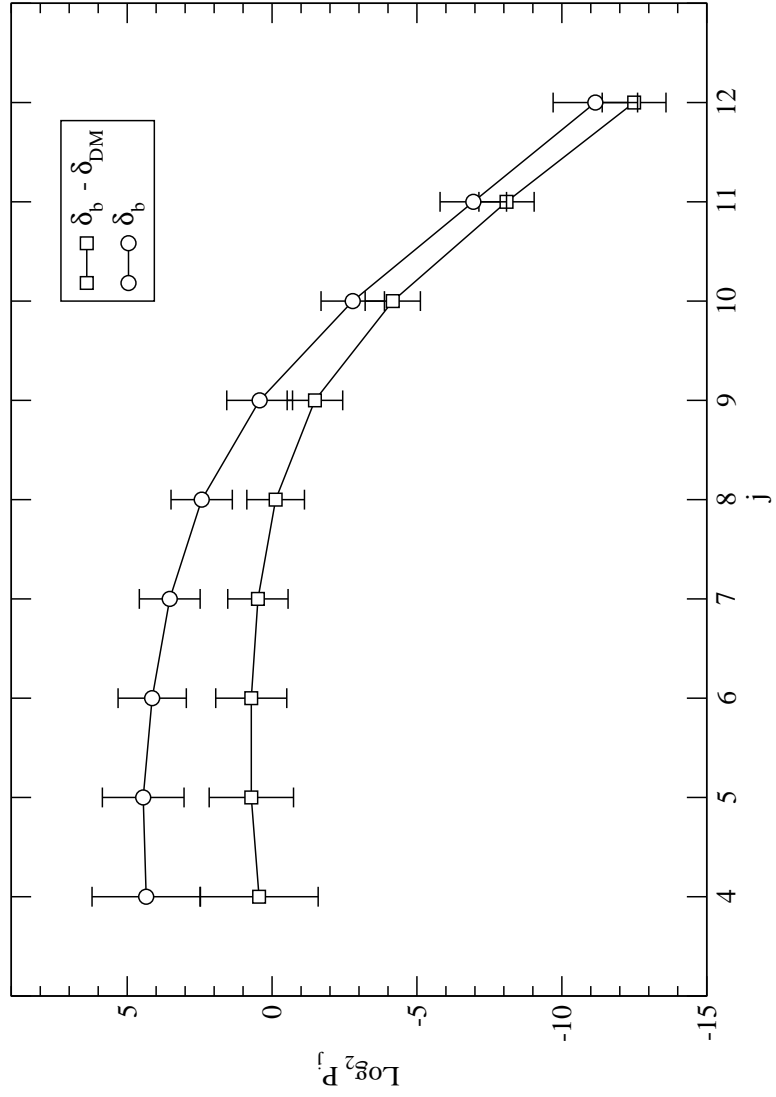}\caption{Power spectra
of $\delta_{b}$ (circle) and $\delta_{b}-\delta_{DM}$ (square) of
the hydro simulation samples for 12 Mpc box. The physical scales
for $j$ is $\Delta v = 2^{13-j} \times 4$ km/s. }
\end{figure}

\begin{figure}
\figurenum{5} \epsscale{0.8} \plotone{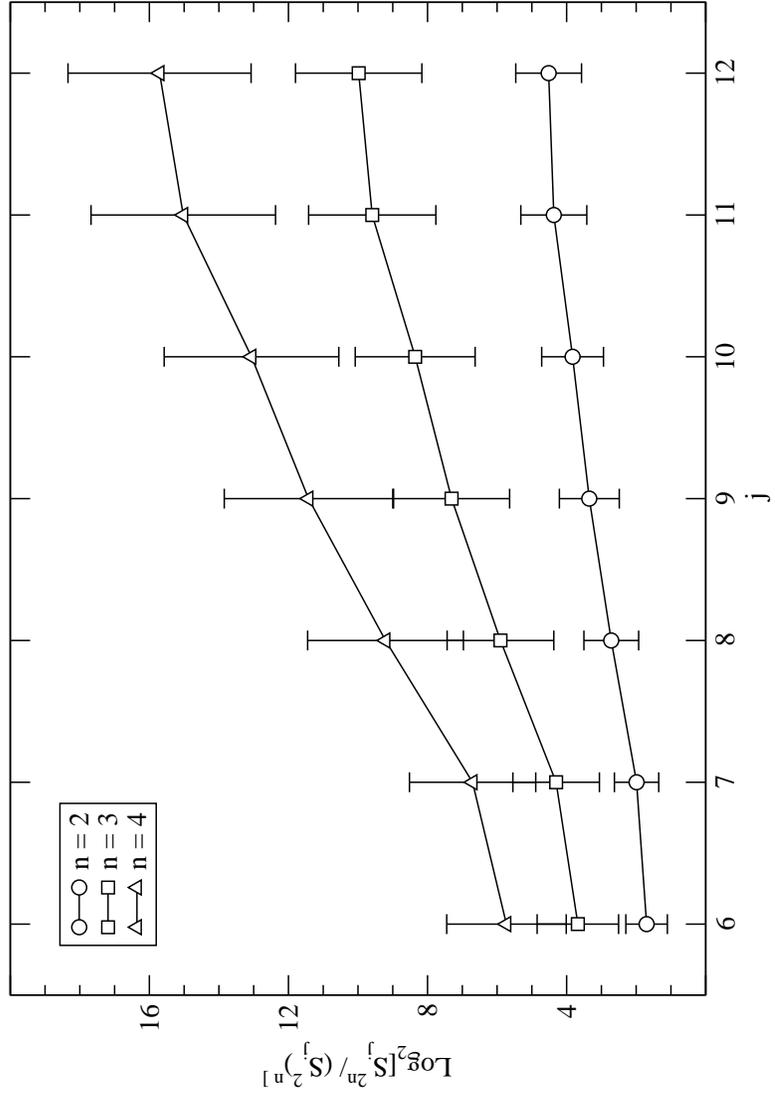}\caption{Structure
functions $\ln_2[S^{2n}_j/(S^2_j)]$ vs. $j$ of the difference
$\delta_{b}-\delta_{DM}$ of the hydro simulation samples in the 12
Mpc box. The physical scale for $j$ is $\Delta v = 2^{13-j} \times
4$ km/s. }
\end{figure}

\begin{figure}
\figurenum{6} \epsscale{0.8} \plotone{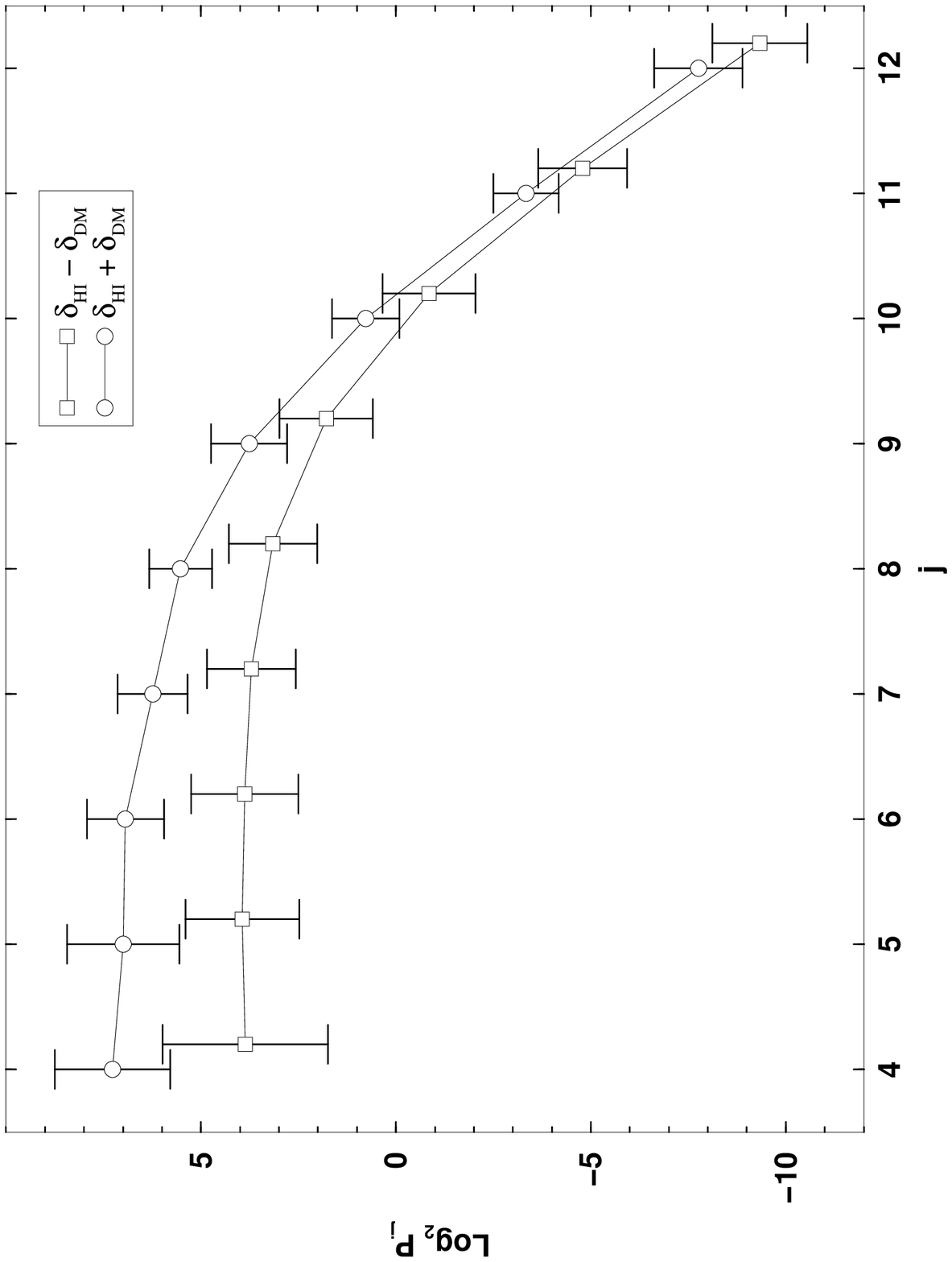}\caption{Power spectra
of $\delta_{HI}+\delta_{DM}$ (circle) and
$\delta_{HI}-\delta_{DM}$ (square) of the 12 Mpc box hydro
simulation samples. The physical scale for $j$ is $\Delta v =
2^{13-j} \times 4$ km/s. For clarity, the power spectrum  of
$\delta_{HI}-\delta_{DM}$ is shifted slightly. }
\end{figure}

\begin{figure}
\figurenum{7} \epsscale{0.8} \plotone{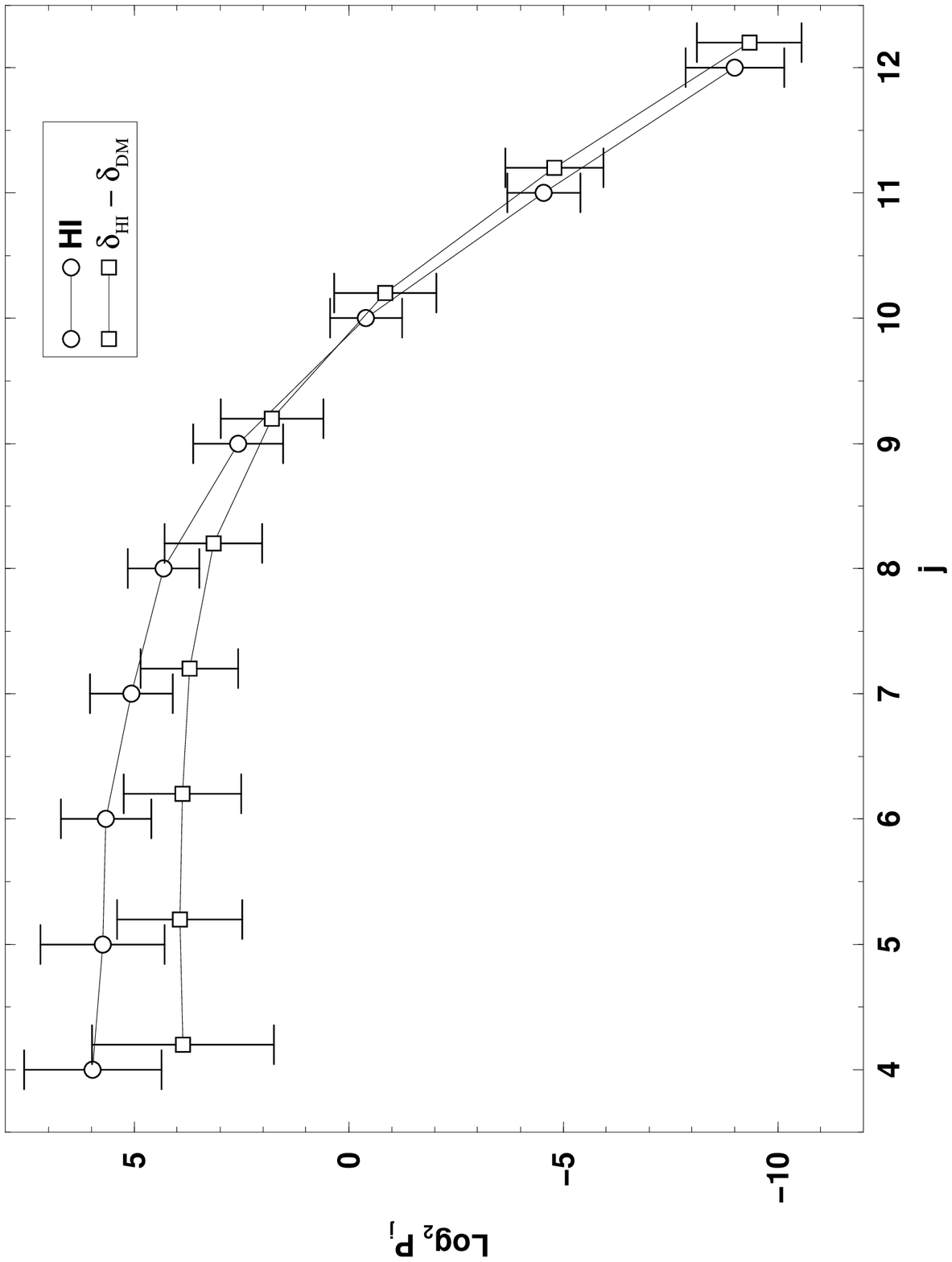} \caption{Power
spectra of $\delta_{HI}$ (circle) and $\delta_{HI}-\delta_{DM}$
(square) of the 12 Mpc box hydro simulation samples. The physical
scales for $j$ is $\Delta v = 2^{13-j} \times 4$ km/s. For
clarity, the power spectrum  of $\delta_{HI}-\delta_{DM}$ is
shifted slightly. }
\end{figure}

\begin{figure}
\figurenum{8} \epsscale{0.8} \plotone{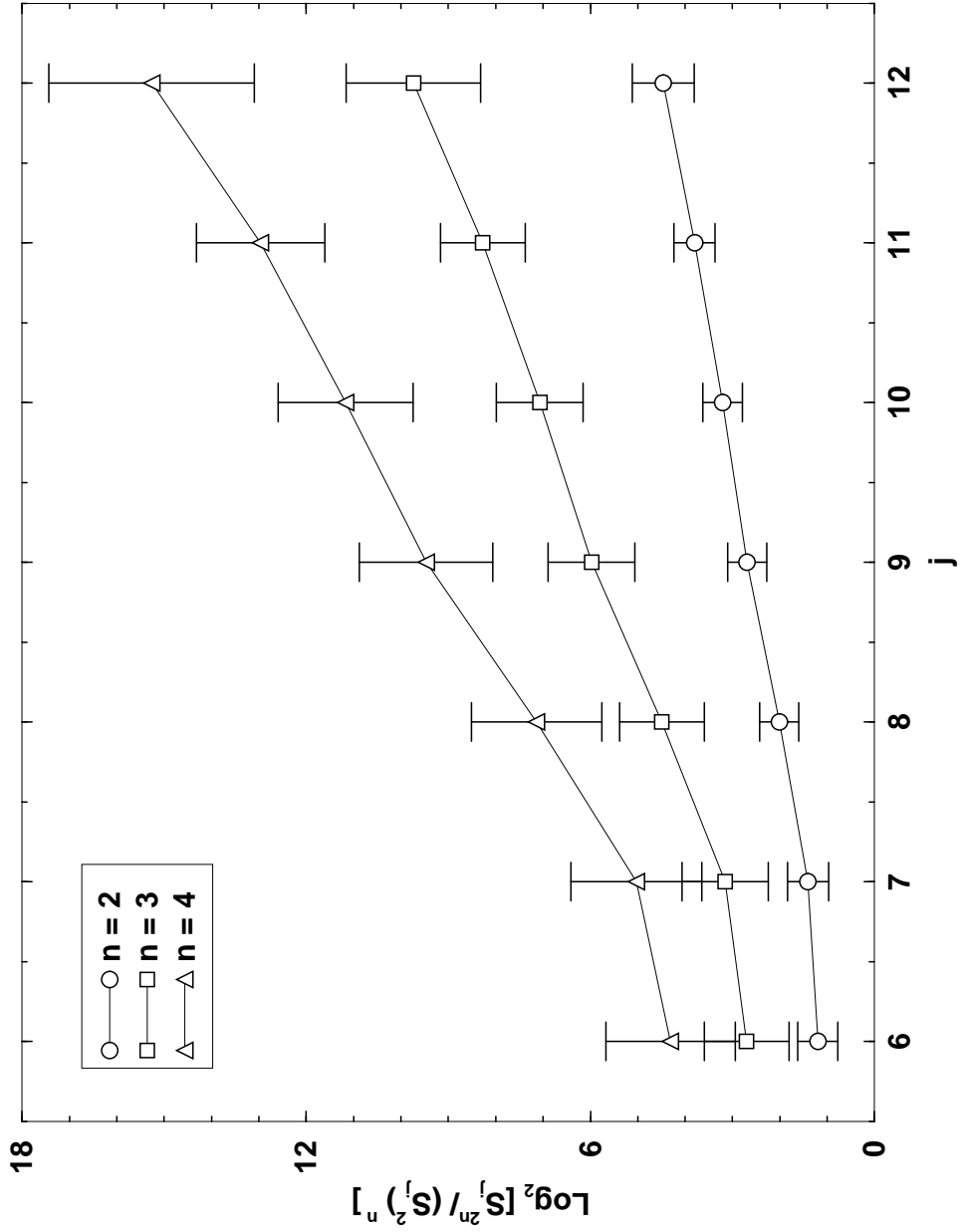} \caption{Structure
functions $\ln_2[S^{2n}_j/(S^2_j)]$ vs. $j$ of the hydro
simulation samples for HS1700+6416 in the 12 Mpc box. The physical
scale for $j$ is $\Delta v = 2^{13-j} \times 4$ km/s. }
\end{figure}

\begin{figure}
\figurenum{9} \epsscale{0.8} \plotone{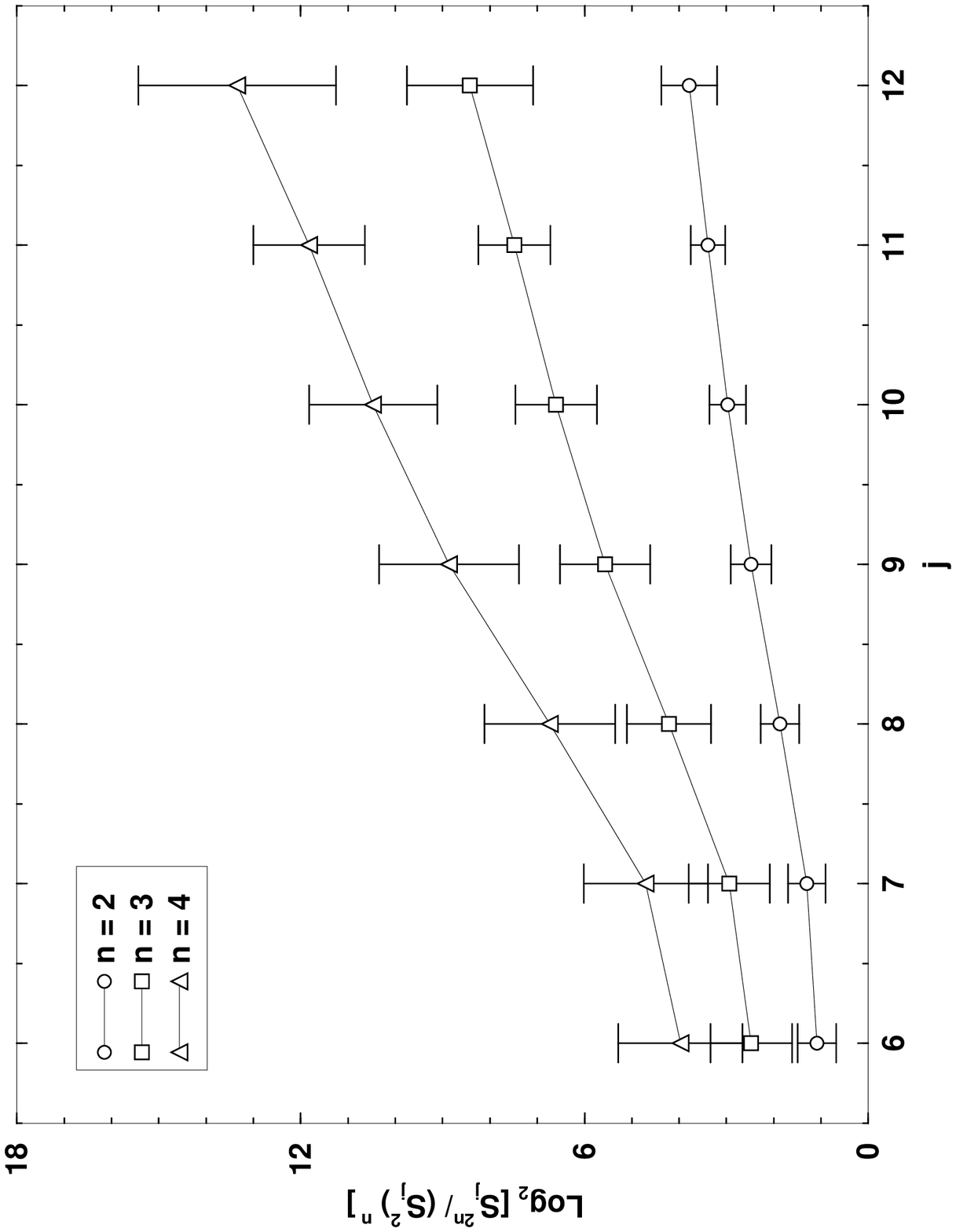} \caption{The same as
Fig. 6, but for q0014+8118. }\
\end{figure}

\begin{figure}
\figurenum{10} \epsscale{0.7} \plotone{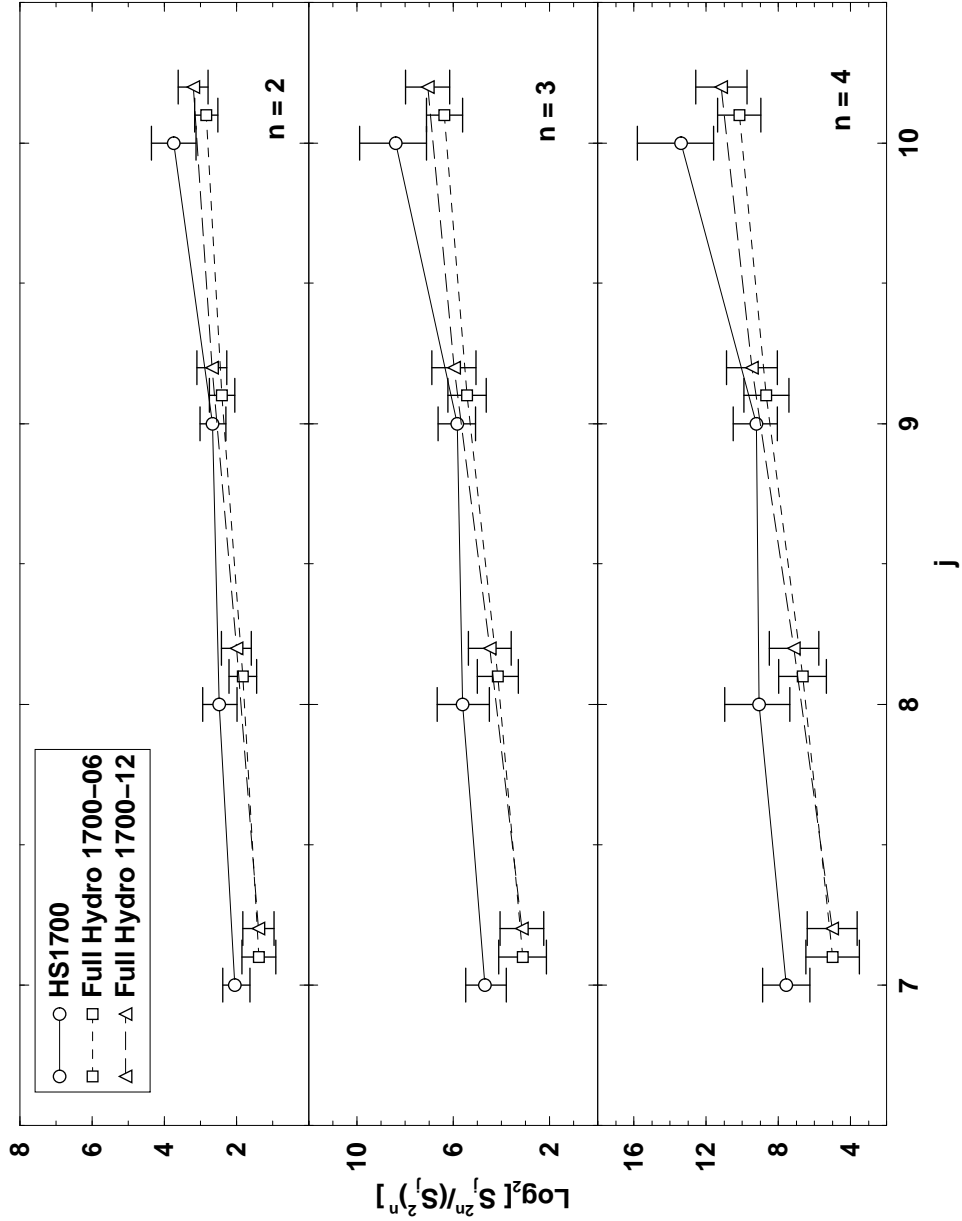} \caption{Structure
function $\ln_2[S^{2n}_j/(S^2_j)]$ vs. $j$ for HS1700+6416 of real
data (circle) and hydro simulation with 12 (triangle) and 6
(square) Mpc box. The physical scale for $j$ is $\Delta v =
2^{13-j} \times 4$ km/s. For clarity, the results of the two
simulation samples are shifted slightly. }
\end{figure}

\begin{figure}
\figurenum{11} \epsscale{0.7} \plotone{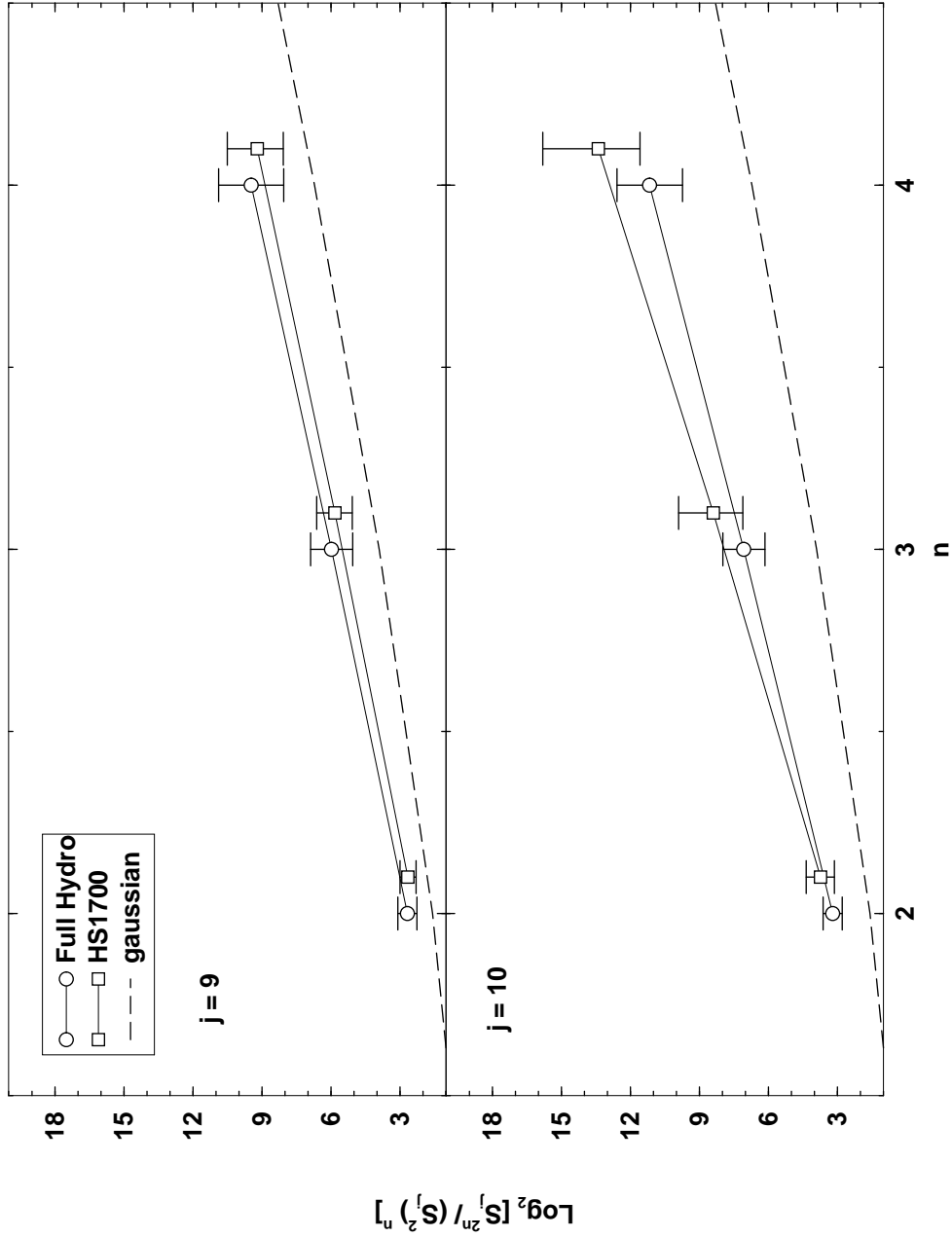} \vspace{1.0cm}\caption{$\ln
S^{2n}_j/(S^{2}_j)^n$ vs. $n$ for simulation sample (circle) and
real data (square) of HS1700+6416, and $j=9$ and 10, or $\Delta v
= 64$ and 32 km/s. The dash line is for a Gaussian field. For
clarify, the results of HS1700 data are shifted a little slightly.
}
\end{figure}

\begin{figure}
\figurenum{12} \epsscale{0.7} \plotone{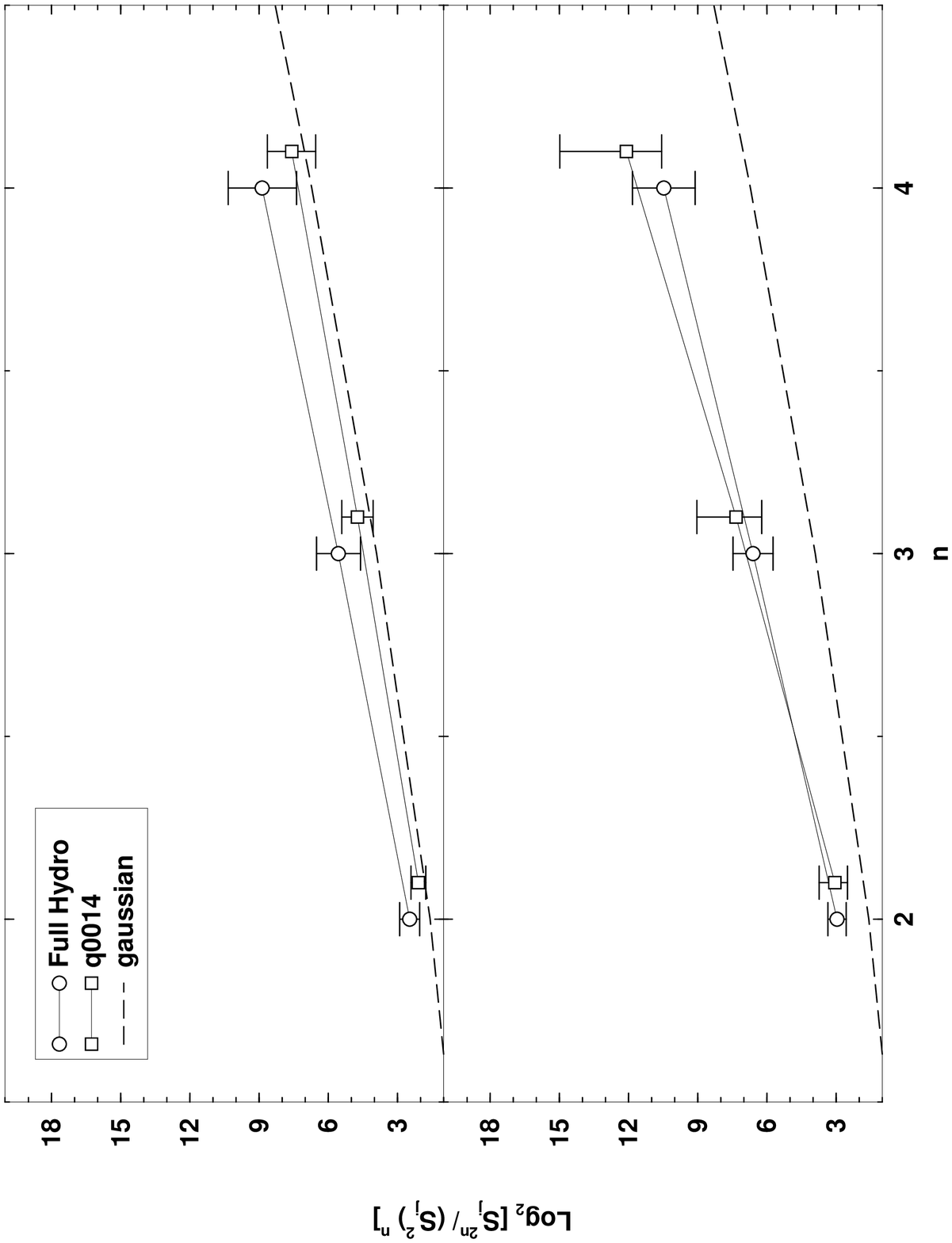}\vspace{1.0cm}\caption{The same
as Fig. 11, but for q0014+8118. }
\end{figure}

\begin{figure}
\figurenum{13} \epsscale{0.8} \plotone{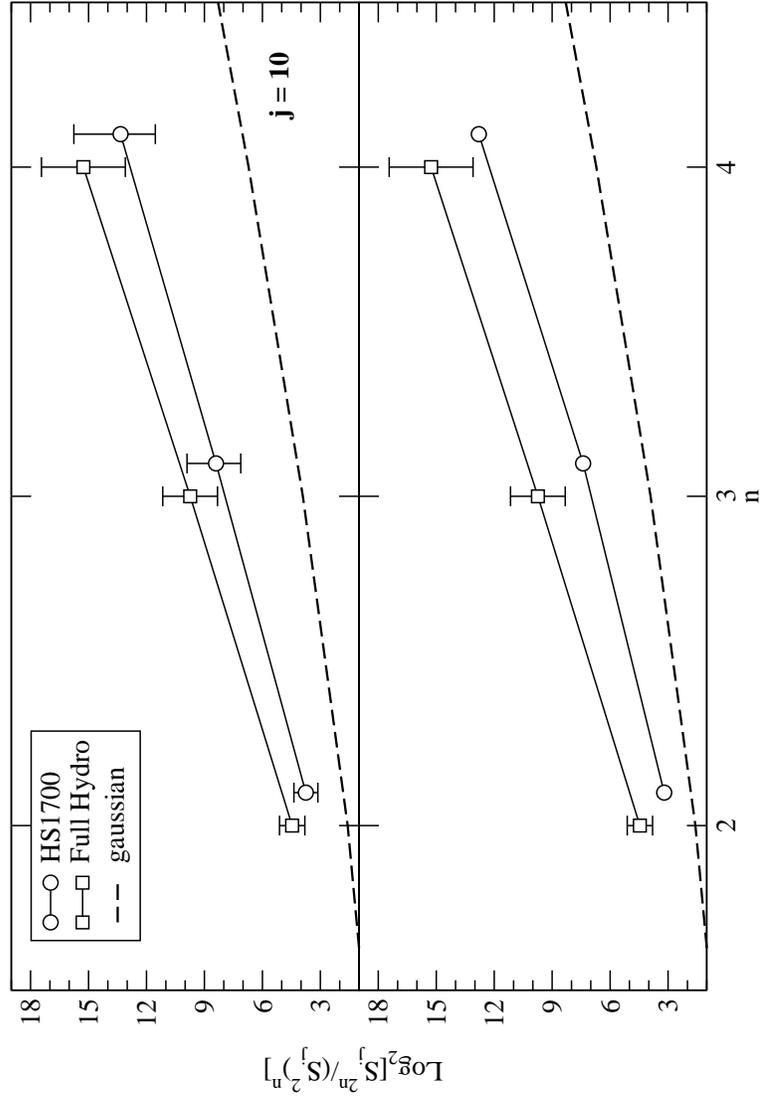}\caption{The same as
Fig. 11 for $j=10$. The top panel is for real data with metal line
removing. The bottom panel is without metal line removing, which
has about the same error bars as the top panel. The dash line is
for a Gaussian field. For clarify, the results of HS1700 data are
shifted a little slightly. }
\end{figure}

\begin{figure}
\figurenum{14} \epsscale{0.8} \plotone{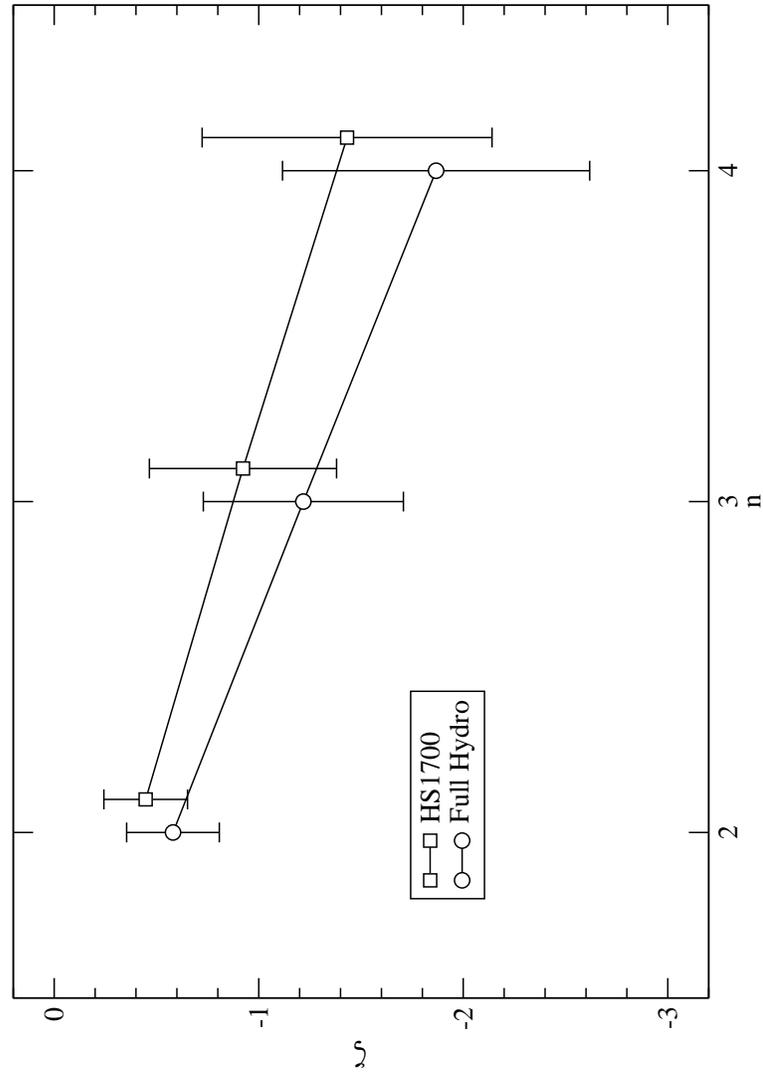}
\caption{Intermittent exponent $\zeta$ vs. $n$ for simulation data
(circle) and real sample (square) of HS1700. For clarity, the
results of HS1700 data are shifted a little bit to right. }
\end{figure}

\begin{figure}
\figurenum{15} \epsscale{0.8} \plotone{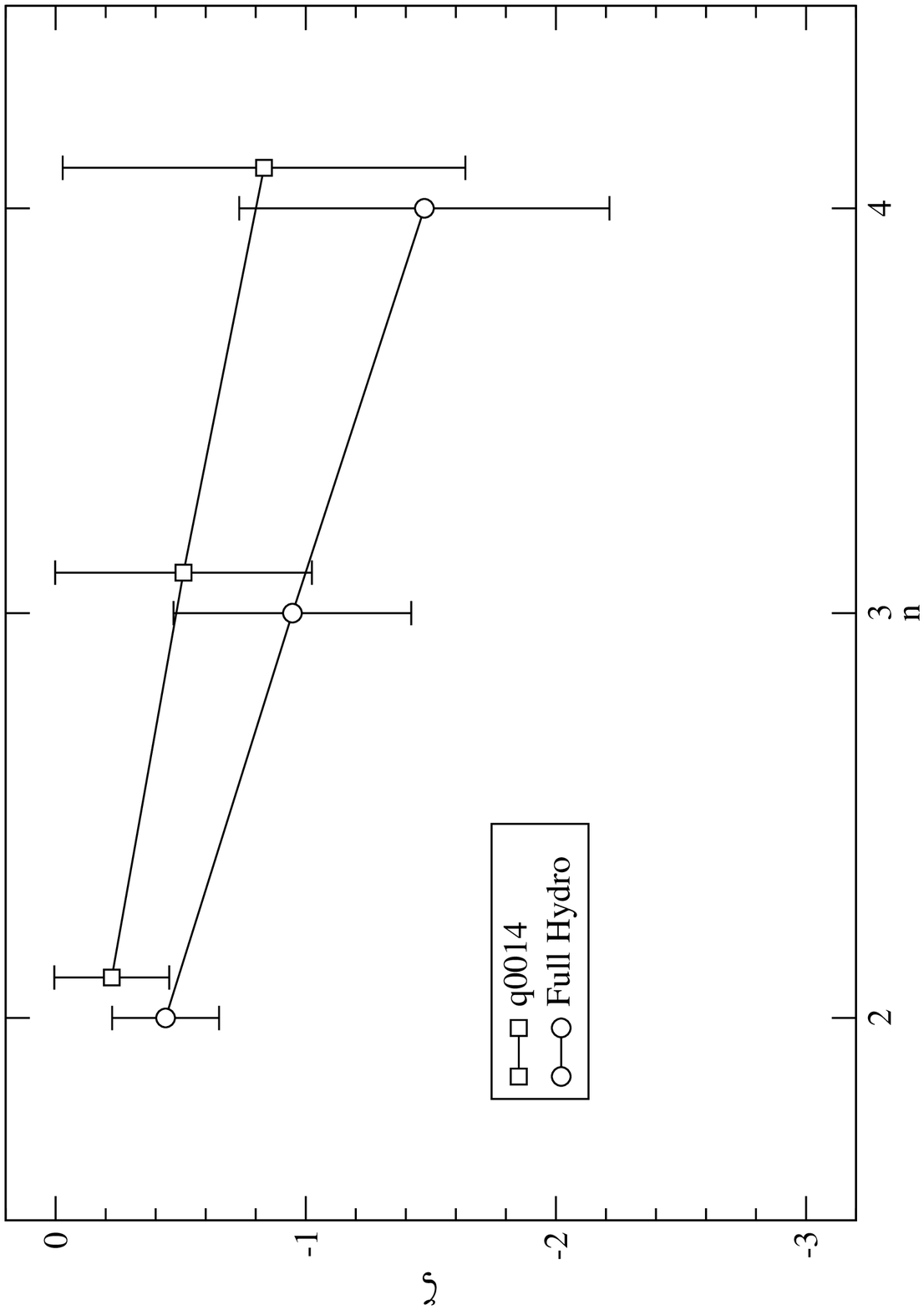} \figcaption{The
same as Fig. 11, but for q0014+8118 }
\end{figure}

\begin{figure}
\figurenum{16} \epsscale{0.8} \plotone{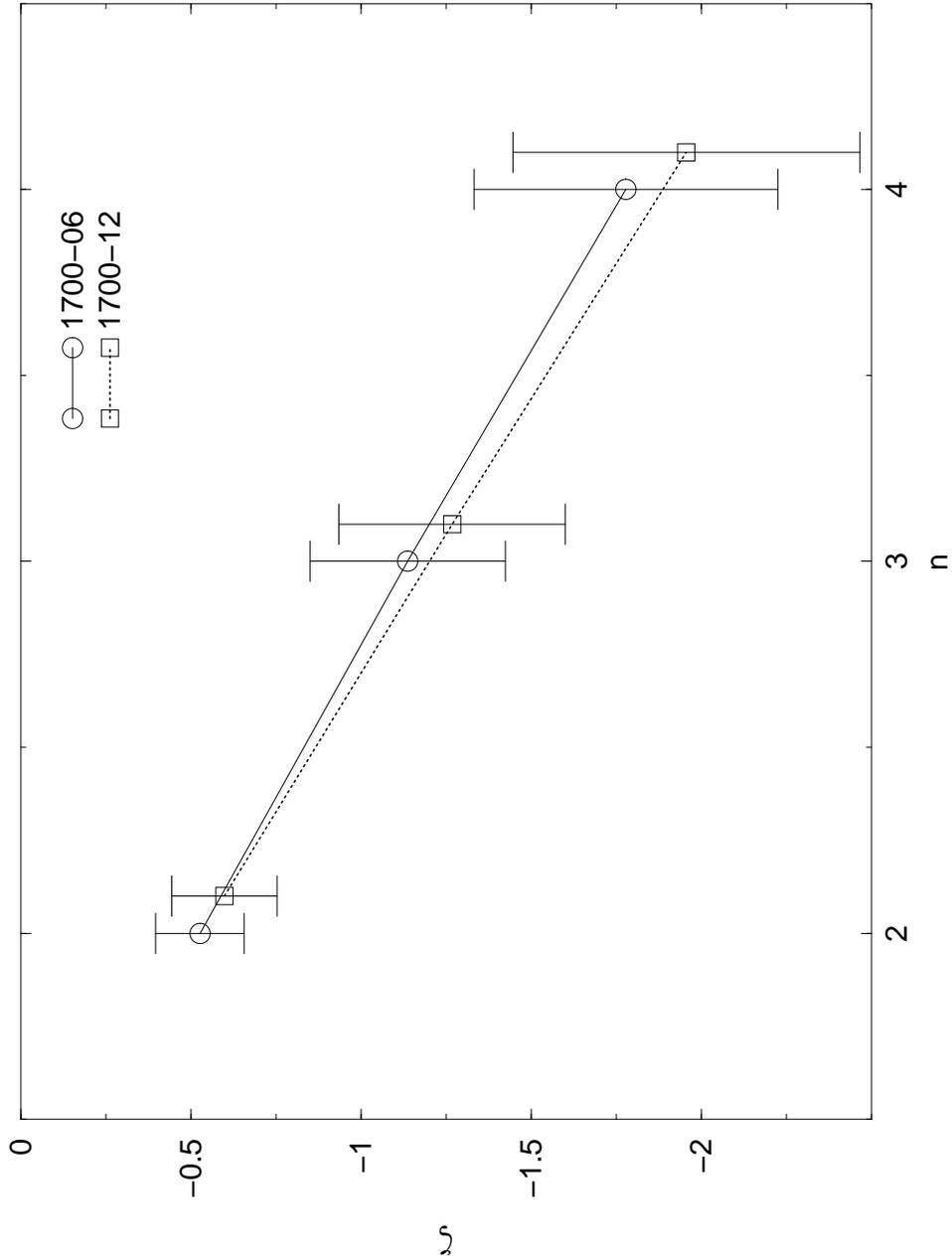}
\caption{Intermittent exponent $\zeta$ vs. $n$ of HS1700 samples
given by hydro simulation in 12 (square) and 6 (circle) Mpc box.
For clarify, the results for 12 Mpc box are shifted a little to
the right. }
\end{figure}

\begin{figure}
\figurenum{17} \epsscale{0.8} \plotone{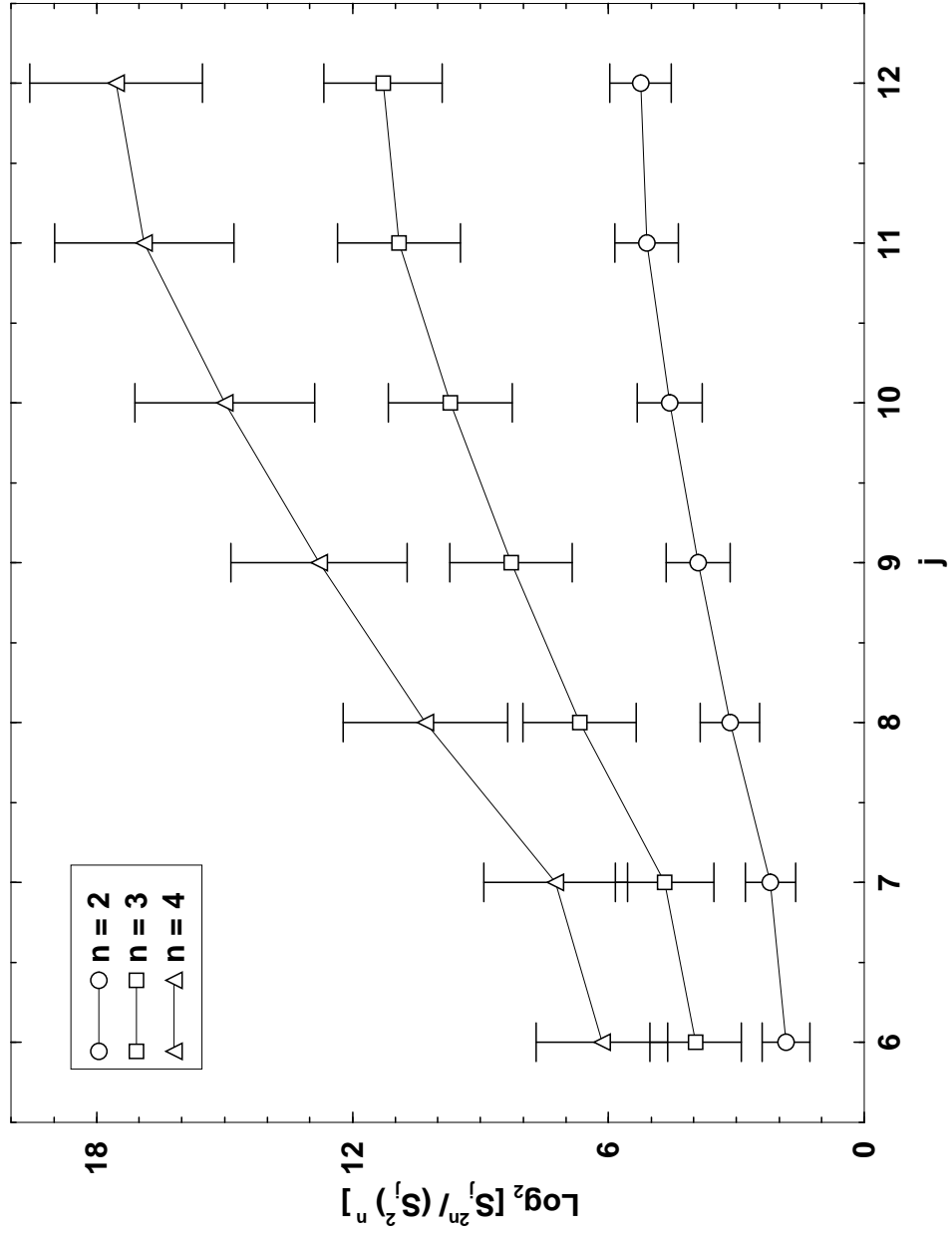}
\figcaption{Structure functions $\ln_2[S^{2n}_j/(S^2_j)]$ vs. $j$
of HI density field $\delta_{HI}$. The physical scale for $j$ is
$\Delta v = 2^{13-j} \times 4$ km/s. }
\end{figure}

\end{document}